\documentclass[conference]{IEEEtran}
\IEEEoverridecommandlockouts
\usepackage{cite}
\usepackage{amsmath,amssymb,amsfonts}
\usepackage{algorithmic}
\usepackage{graphicx}
\usepackage{textcomp}

\usepackage{listings}
\usepackage{graphicx}
\usepackage{comment}
\usepackage{color}
\usepackage{subfigure}
\usepackage{epstopdf}
\usepackage{url}
\usepackage{kantlipsum}
\usepackage{caption}

\definecolor{dkgreen}{rgb}{0,0.6,0}
\definecolor{gray}{rgb}{0.5,0.5,0.5}
\definecolor{mauve}{rgb}{0.58,0,0.82}

\lstdefinestyle{myScalastyle}{
	frame=tb,
	language=scala,
	aboveskip=3mm,
	belowskip=3mm,
	showstringspaces=false,
	columns=flexible,
	basicstyle={\small},
	numbers=none,
	numberstyle=\tiny\color{gray},
	keywordstyle=\color{blue},
	commentstyle=\color{dkgreen},
	stringstyle=\color{mauve},
	breaklines=true,
	breakatwhitespace=true,
	tabsize=3,
	captionpos=b,
}

\def\BibTeX{{\rm B\kern-.05em{\sc i\kern-.025em b}\kern-.08em
    T\kern-.1667em\lower.7ex\hbox{E}\kern-.125emX}}

\newif\ifdblind
\dblindfalse
\ifdblind
\newcommand{\mesh}{HPS}
\else
\newcommand{\mesh}{MESH} 
\fi

\begin{document}

\title{MESH: A Flexible Distributed Hypergraph Processing System}

\author{\IEEEauthorblockN{Benjamin Heintz}
\IEEEauthorblockA{\textit{University of Minnesota} \\
	Minneapolis, MN \\
	heintz@umn.edu}
\\
\IEEEauthorblockN{Gaurav Khandelwal}
\IEEEauthorblockA{\textit{University of Minnesota} \\
	Minneapolis, MN \\
	khand052@umn.edu}
\and
\IEEEauthorblockN{Rankyung Hong}
\IEEEauthorblockA{\textit{University of Minnesota} \\
	Minneapolis, MN \\
	hongx293@umn.edu}
\\
\IEEEauthorblockN{Corey Tesdahl}
\IEEEauthorblockA{\textit{University of Minnesota} \\
	Minneapolis, MN \\
	tesd0005@umn.edu}
\and
\IEEEauthorblockN{Shivangi Singh}
\IEEEauthorblockA{\textit{University of Minnesota} \\
	Minneapolis, MN \\
	singh486@umn.edu}
\\
\IEEEauthorblockN{Abhishek Chandra}
\IEEEauthorblockA{\textit{University of Minnesota} \\
	Minneapolis, MN \\
	chandra@umn.edu}
}

\maketitle

\begin{abstract}
  With the rapid growth of large online social networks, the ability to analyze
  large-scale social structure and behavior has become critically important,
  and this has led to the development of several scalable graph processing
  systems.
  In reality, however, social interaction takes place not only between pairs of
  individuals as in the graph model, but rather in the context of multi-user
  groups.
  Research has shown that such group dynamics can be better modeled through a
  more general \emph{hypergraph} model, resulting in the need to build scalable
  hypergraph processing systems. 
  In this paper, we present \mesh{}, a flexible distributed framework for scalable
  hypergraph processing. \mesh{} provides an easy-to-use and expressive application programming interface
  that naturally extends the ``think like a vertex'' model common to many
  popular graph processing systems.
  Our framework provides a flexible implementation based on an underlying graph processing system, and enables different
  design choices for the key implementation issues of partitioning a hypergraph representation. 
  We implement \mesh{} on top of the popular GraphX graph processing framework in
  Apache Spark. Using a variety of real datasets and experiments conducted on a local 8-node cluster as well as a 65-node Amazon AWS testbed, we demonstrate that \mesh{} provides flexibility based on data
  and application characteristics, as well as scalability with cluster size.
We further show that it is competitive in performance to HyperX, another
  hypergraph processing system based on Spark, 
while providing a much simpler implementation (requiring about 5X fewer lines of code), thus
showing that simplicity and flexibility need not come at the cost of performance.
\end{abstract}

\section{Introduction}
\label{sec:introduction}

The advent of online social networks and communities such as Facebook and
Twitter has led to unprecedented growth in user interactions (such as
``likes'', comments, photo sharing, and tweets), and collaborative activities
(such as document editing and shared quests in multi-player games).
This has resulted in massive amounts of rich data that can be analyzed to
better understand user behavior, information flow, and social dynamics.
The traditional way to study social networks is by modeling them as {\em
graphs}, where each vertex represents an entity (e.g., a user) and each edge
represents the relation or interaction between two entities (e.g., friendship). 
Myriad graph analytics
frameworks~\cite{gonzalez2014,malewicz2010,nguyen2013}
have been introduced to scale out the computation on massive graphs comprising
millions or billions of vertices and edges.

While graph analytics has enabled a better understanding of social interactions
between individuals, there is a growing interest~\cite{lazer2006} in studying
{\em groups} of individuals as entities on their own.
A group is an underlying basis for many social interactions and collaborations,
such as users on Facebook commenting on an event of common interest, or a team
of programmers collaborating on a software project.
In these cases, individuals interact in the context of the overall group, and
not simply in pairs.
Further, the dynamics of many such systems may also be driven through
group-level events, such as users joining or leaving groups, or finding others
based on group characteristics (e.g., common interest). 

Since such group-based phenomena involve multi-user interactions, it has been
shown that many natural phenomena can be better modeled using {\em hypergraphs} than by using graphs~\cite{estrada2005} ranging from large-scale social graphs~\cite{sharma2017weighted} to
disease-gene networks~\cite{gallagher2014}. As a result, there is a growing need for scalable hypergraph processing systems that 
can enable easy implementation and efficient execution of such algorithms on real-world data.

\begin{figure}
  \centering
  \subfigure[Graph]{
    \includegraphics[width=0.40\columnwidth]{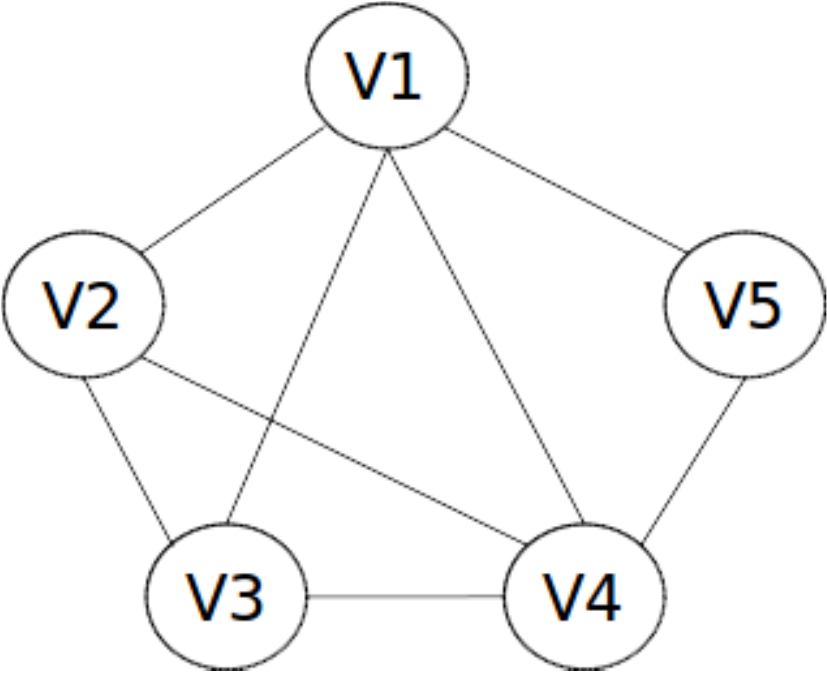}
    \label{fig:three_authors_graph}
  }
  \subfigure[Hypergraph]{
    \includegraphics[width=0.48\columnwidth]{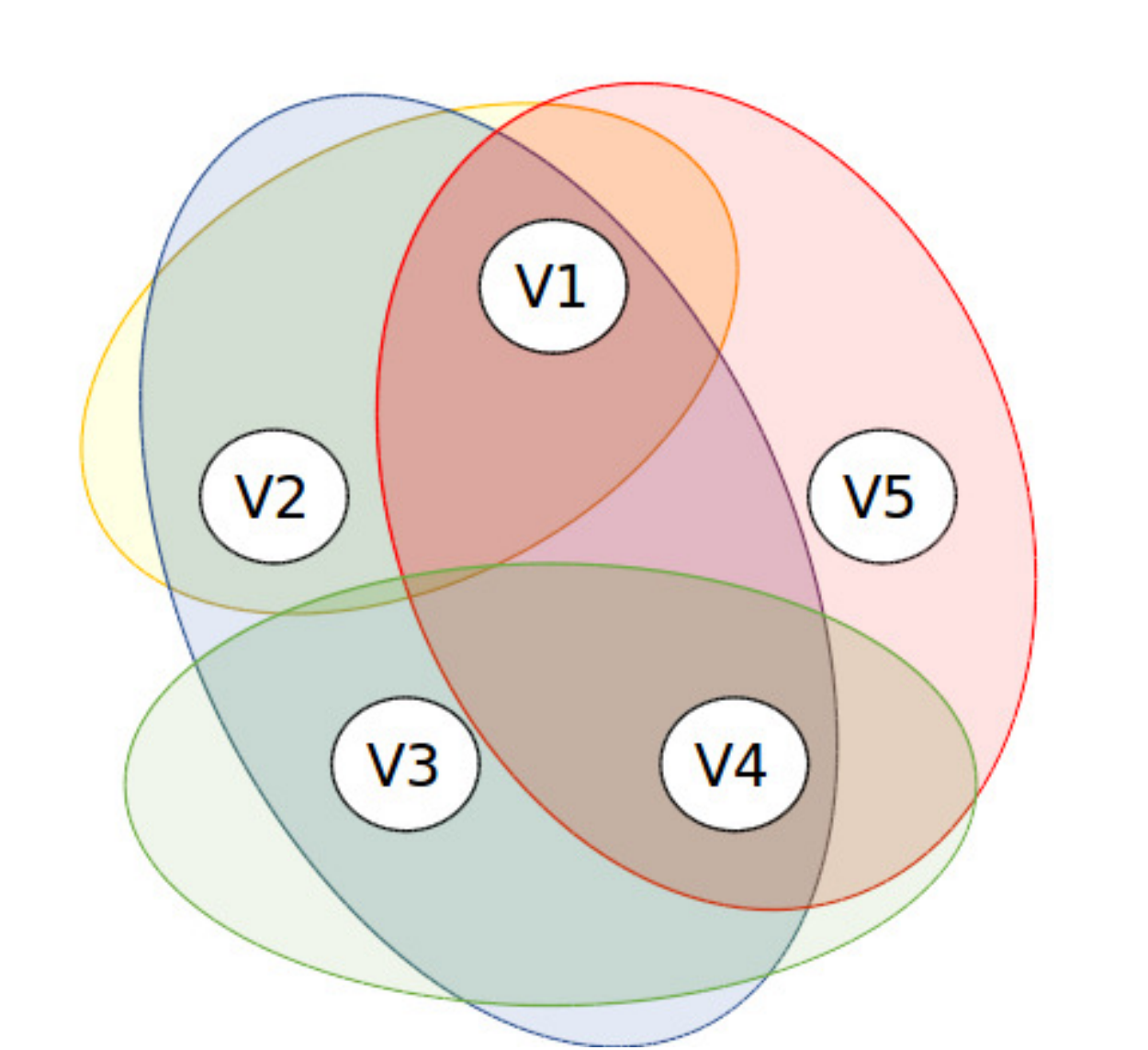}
    \label{fig:three_authors_hypergraph}
  }
  \vspace{-0.25em}
  \caption{A hypergraph can model groups unambiguously compared to a simple graph.
    Here, we have four groups ($\{v_1,v_2\}, \{v_1,v_2,v_3,v_4\}, \{v_1,v_4,v_5\}, \{v_3,v_4\}$).}
  \vspace{-1em}
  \label{fig:graph-vs-hgraph}
\end{figure}

Formally, a {\em hypergraph} is a generalization of a graph\footnote{In this
paper, we use ``graph'' to refer to a traditional dyadic graph.}, and is
defined as a tuple $H = (V,E)$, where $V$ is the set of entities, called {\em
vertices}, in the network, and $E$ is the set of subsets of $V$, called {\em
hyperedges}, representing relations between one or more
entities~\cite{berge1976} (as opposed to exactly two in a graph).
Figure~\ref{fig:graph-vs-hgraph} illustrates the difference between a graph and a hypergraph. This figure shows a 5-vertex network, consisting of four groups ($\{v_1,v_2\}, \{v_1,v_2,v_3,v_4\}, \{v_1,v_4,v_5\}, \{v_3,v_4\}$).
As can be seen from the figure, a graph can only capture binary relations (e.g., $\{v_1,v_2\},
\{v_3,v_4\}$, etc.), some of which may correspond to distinct overlapping groups (e.g., $\{v_3,v_4\}$ belongs to two distinct groups).
On the other hand, 
a hypergraph can model all the groups unambiguously compared to a graph.

From a system standpoint, a hypergraph processing system must satisfy 
several design goals.
First, for easy adoption by users, a hypergraph processing system must provide an interface that
is \textit{expressive and easy-to-use} by application programmers.
Second is the ability to handle data at different scales,
ranging from small hypergraphs to massive ones (with millions or billions of vertices and hyperedges). 
As a result, similar to a graph processing system, a hypergraph processing
system must be \textit{scalable}, both in terms of memory and storage
utilization, as well as by enabling distributed
computation across multiple CPUs and nodes for increased parallelism as needed.
Third, it must be \textit{flexible} in order to perform well in the face of diverse
application and data characteristics.
Finally, any novel design for a hypergraph processing system should strive for
\textit{ease of implementation}, as this allows faster development,
enhancement, and maintenance.

From a high level, there are two main approaches to building a hypergraph
processing system.
One approach is to build a specialized system for hypergraph processing  from scratch (e.g., HyperX~\cite{huang2015}).
While this approach has the benefit of allowing hypergraph-specific
optimizations at a lower level, it can be limited in terms of its flexibility and
may require a sophisticated implementation effort.
A different approach is to overlay a hypergraph processing system {\em on top of} an existing graph
processing system. This approach can leverage many
mechanisms and optimizations already available in existing mature graph
processing systems, and hence, can be simpler to implement, and can
provide flexibility in terms of design choices. 
We take this approach, exploring the issues and tradeoffs involved therein to show its efficacy. 

In this paper, we present 
\mesh{}%
\ifdblind
\footnote{We refer to our system by a generic name \mesh{} to preserve anonymity due to double-blind requirements.} (Hypergraph Processing System),
\else
\footnote{{\bf M}innesota {\bf E}ngine for {\bf S}calable {\bf H}ypergraph analysis}, 
\fi
a distributed hypergraph processing system based on 
a graph processing framework.
We use our system to explore the key challenges on how to partition a hypergraph representation
to allow efficient distributed computation in implementing a
hypergraph processing system on top of a graph processing system.
For our implementation, we choose the GraphX
framework~\cite{gonzalez2014} in Apache Spark~\cite{zaharia2012} due to its popularity and mature software eco-system,
though we expect our ideas to be applicable or extensible to
other graph processing frameworks as well.

\subsection{Research Contributions}

  \begin{itemize}
  \item We present \mesh{}, a distributed hypergraph processing system designed
    for scalable hypergraph processing, based on a graph processing framework.

  \item We present an expressive API for hypergraph processing, which extends the
    popular ``think like a vertex'' programming model~\cite{malewicz2010} by
    treating hyperedges as first-class computational objects with their own
    state and behavior (Section~\ref{sec:api}).

  \item We explore the impact of the key design question in building a hypergraph processing system: how to partition hypergraph representations for distributed computation. 
We present multiple hypergraph partitioning algorithms and show how to map them to graph partitioning algorithms (Section~\ref{sec:implementation}).

  \item We implement a \mesh{} prototype%
\footnote{We have released the source code for our implementation,
but do not reveal the repository for double-blind reasons.}
on top of the GraphX graph processing
    system built on Apache Spark (Section~\ref{sec:evaluation}).
    Using this prototype and a number of real datasets and algorithms, 
we conduct experiments on a local 8-node cluster as well as a 65-node Amazon AWS testbed (Section~\ref{sec:evaluation}).
We experimentally demonstrate that \mesh{} provides the flexibility to make 
design choices based on data and application characteristics, and achieves scalability with cluster size.
We further show that our \mesh{} implementation is competitive in performance to HyperX~\cite{huang2015}  hypergraph processing system,
while providing a much simpler implementation (requiring about 5X fewer lines of code), thus
showing that simplicity and flexibility need not come at the cost of performance.
\end{itemize}

   % label: sec:introduction
\section{\mesh{} Overview}
\label{sec:overview}

\subsection{Design Goals}

\noindent{\bf Expressiveness \& Ease of Use:}
    Hypergraph algorithms are fundamentally more general than graph algorithms.
    Many hypergraph algorithms treat hyperedges as first-class entities
    on par with vertices.
    A hypergraph processing system should therefore be \emph{expressive} enough
    to allow hyperedges to have attributes and computational functions just as
    vertices do.
    It is critical that these attributes and functions be as general for
    hyperedges as they are for vertices.
    In addition to this expressiveness,
    a hypergraph system should also provide \emph{ease of use}, enabling
    application developers to easily write a diverse variety of hypergraph
    applications. %without unnecessary complication.

\noindent{\bf Scalability:}
Many real-world datasets range in size from small to massive, comprising
    millions or billions of vertices and hyperedges.
    Similar to popular graph processing systems, hypergraph processing systems
    must be designed to scale to massive inputs, and they must allow
    distributed processing over multiple machines, while efficiently processing small datasets as well.

\noindent{\bf Flexibility:}
    A hypergraph processing system must answer two key questions of
    how to represent hypergraphs, and how to partition this representation
    for distributed computation.
    As we show in Section~\ref{sec:implementation}, the right answer to these
    questions depends on many factors related to the input dataset and
    algorithm characteristics.
    A hypergraph processing system must therefore be \emph{flexible}, allowing
    the appropriate answers for these questions to be made at runtime based on
    data and application characteristics.

\noindent{\bf Ease of Implementation:}
A hypergraph processing system should be
    designed to simplify implementation as much as practical.
    This not only allows for faster development with fewer defects, but it
    allows the system to evolve more rapidly as it gains adoption.
    This is especially important as hypergraph processing is a novel area, where
    applications and systems will need to evolve rapidly in tandem. % in the    coming years.

Existing graph processing systems such as Pregel~\cite{malewicz2010},
PowerGraph~\cite{gonzalez2012}, and GraphX~\cite{gonzalez2014} provide the
foundation for scalability.
They also provide a useful pattern we can follow to achieve programmability,
namely the ``think like a vertex'' programming model, where graph processing
applications
are expressed in terms of vertex-level programs that iteratively receive messages
from their neighbors, update their state, and send message to their neighbors.
%in a series of ``supersteps''.
As we will show, however, these existing systems lack the flexibility required
to handle diverse hypergraph applications and data.

\subsection{\mesh{} Hypergraph Processing System}

\begin{figure}
  \centering
  \includegraphics[width=0.50\columnwidth]{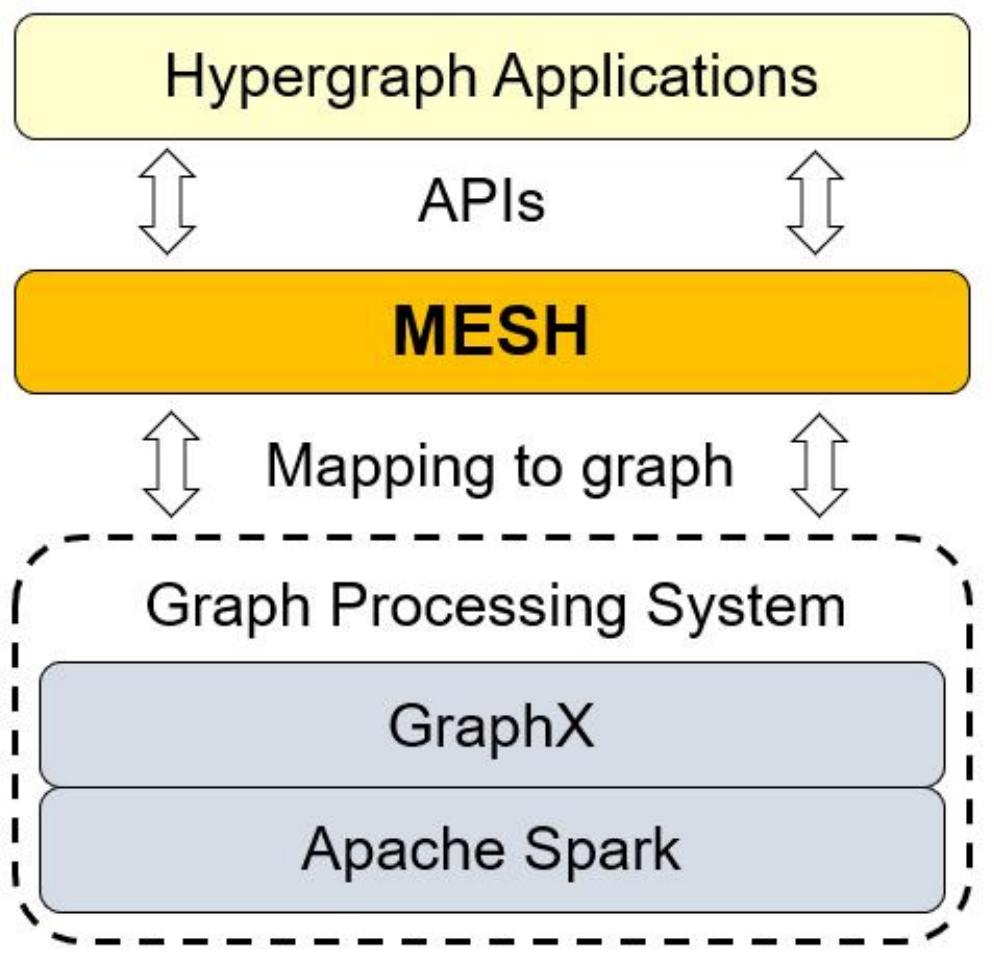}
  \caption{\mesh{} is implemented on top of a graph processing engine 
and provides an expressive
    and easy-to-use API to hypergraph applications.
  \label{fig:mesh}}
\end{figure}

In order to meet the requirements of scalability and ease of implementation, 
we focus on implementing our hypergraph processing system, called
\mesh{},
on top of an existing graph processing system
rather than from scratch.
We assume that the underlying graph processing framework provides us with
a graph representation consisting of vertices and edges, a graph partitioning framework to partition the input data across multiple machines, and a distributed execution framework that supports computation and communication across multiple machines, along with some fault tolerance mechanisms.
For our implementation, we choose the GraphX
framework~\cite{gonzalez2014} in Apache Spark~\cite{zaharia2012}.
As Figure~\ref{fig:mesh} shows, \mesh{} is positioned as a middleware layer between
hypergraph applications and GraphX.

Given such a system architecture, 
we explore two key research challenges throughout this paper: developing an
expressive and easy-to-use API for enabling diverse hypergraph algorithms (Section~\ref{sec:api}), and implementing this
API on top of an existing graph processing system
(Section~\ref{sec:implementation}).
       % label: sec:overview
\section{Application Programming Interface}
\label{sec:api}

In this section, we first discuss the features of hypergraph algorithms and then 
present the HPS API that can enable expressing such algorithms easily.

\subsection{Hypergraph Algorithms}
\label{sec:algos}
Many hypergraph algorithms can be viewed as generalizations of
corresponding graph algorithms, but they can have richer attributes and
computations, 
particularly those defined for hyperedges in addition to vertices.
We examine some example hypergraph algorithms below to illustrate these aspects.

\subsubsection{PageRank}
\label{subsec:pr}
Consider PageRank~\cite{page1999}, a widely used algorithm
in graph analytics to determine the relative importance of different vertices
in a graph.
It is used in a variety of applications, such as search, link prediction, and
recommendation systems.

We can extend PageRank to the hypergraph context in many ways.
The most straightforward extension is to compute the PageRank for vertices
based on their membership in different hyperedges.
In a social context, this would correspond to determining the importance of a
user based on her group memberships (e.g., a user might be considered more
important if she is part of an exclusive club). % exclusive club --> important group

At the same time, it is possible to compute the PageRank for {\em hyperedges} based
on the vertices they contain.
This corresponds to estimating the importance of groups based on their members
(e.g., a group with Fortune 500 CEOs is likely to be highly important). 
This extension also illustrates the fact that hyperedges can be considered
first-class entities associated with similar state and computational functions
as vertices in typical graph computation.

This elevation of hyperedges to first-class status enables further extensions
to PageRank: we can compute additional attributes for hyperedges using
arbitrary functions of their member vertices.
For example, we can use an entropy function to determine the uniformity of each
hyperedge; i.e., the extent to which its members contribute equally to its
importance.

\subsubsection{Label Propagation}
\label{subsec:lp}
Consider a Label Propagation algorithm~\cite{huang2015,raghavan2007}, which determines the community
structure of a hypergraph.
Here, in addition to identifying the community to which each vertex belongs, we
may also assign to each {\em hyperedge} the community to which it belongs.
Such an algorithm proceeds by iteratively passing messages from vertices to
hyperedges and back.
At each step along the way, the solution is refined as vertices and hyperedges
update their attributes to record the community to which they belong.

\subsubsection{Shortest Paths}
\label{subsec:sp}
Consider the Single Source Shortest Paths algorithm that computes the shortest paths
from a source vertex to all other vertices in the network.
In the hypergraph context, a path would be defined in terms of the hyperedges
that are traversed from the source to each destination, and the path length would
depend on the number of hyperedges along the path as well as any weights assigned to them.
As an example, this can allow us to compute the degree of separation between two users in terms of the group structure of a social network.
Conversely, one can also compute shortest paths between hyperedges (e.g., to identify how far two groups are in terms of the connectivity of their users).

Along these same lines, hypergraph extensions can be derived for many popular
graph algorithms, such as connected components, %shortest paths, 
centrality estimation~\cite{roy2015}, and more.
The key to this expressiveness is the elevation of hyperedges to first-class
status.

\subsection{Core API}
To make \mesh{} easy to use, its API builds upon programmers' existing
familiarity with the ``think like a vertex'' model~\cite{malewicz2010}, by providing a ``think like
a vertex \emph{or hyperedge}'' model.
\mesh{} provides an iterative computational model similar to Pregel,
but with the introduction of hyperedges as first-class entities with their own
computational behavior and state.
In this model, computation proceeds iteratively in a series of alternating ``supersteps''
(alternating between vertex and hyperedge computation).
Within a superstep, vertices
(resp., hyperedges) update their state and compute new messages, which are
delivered to their incident hyperedges (resp., vertices).

%---------- Approximate column width for IPDPS -----------|
\begin{lstlisting}[
style=myScalastyle,
caption={Key abstractions from our hypergraph API (expressed in Scala).},
label={listing:api},
basicstyle={\small},
emph={HyperGraph, Program, MessageCombiner, Procedure, Context},
emphstyle={\textbf},
tabsize=2
]
trait HyperGraph[VD, HED] {
	def compute[ToHE, ToV](
		maxIters: Int,
		initialMsg: ToV,
		vProgram: Program[VD, ToV, ToHE],
		heProgram: Program[HED, ToHE, ToV])
		: HyperGraph[VD, HED]
	}
	object HyperGraph {
		trait Program[Attr, InMsg, OutMsg] {
		def messageCombiner: MessageCombiner[OutMsg]
		def procedure: Procedure[Attr, InMsg, OutMsg]
	}
	type MessageCombiner[Msg] = (Msg, Msg) => Msg
	type Procedure[Attr, InMsg, OutMsg] =
		(Step, NodeId, Attr, InMsg, Context[Attr, OutMsg]) => Unit
	trait Context[Attr, OutMsg] {
		def become(attr: Attr): Unit
		def send(msgF: NodeId => OutMsg, to: Dst): Unit
		def broadcast(msg: OutMsg): Unit = send(msg, All)
	}
}
\end{lstlisting}

Listing~\ref{listing:api} shows the core of the \mesh{} API\footnote{We show Scala code for our API/algorithms. Scala
  \texttt{trait}s are analogous to Java \texttt{interface}s, and the
  \texttt{object} keyword here is used to define a module namespace.}.
The key abstraction is the \textbf{\lstinline{HyperGraph}}, which is
parameterized on the vertex and hyperedge attribute data types.
Similar to the GraphX \textbf{\lstinline{Graph}} interface, the
\textbf{\lstinline{HyperGraph}} provides methods (not shown) such as
\textbf{\lstinline{vertices}} and \textbf{\lstinline{hyperEdges}} for accessing
vertex and hyperedge attributes, \textbf{\lstinline{mapVertices}} and
\textbf{\lstinline{mapHyperEdges}} for transforming the hypergraph,
\textbf{\lstinline{subHyperGraph}} for computing a subhypergraph based on
user-defined predicate functions, and so on.

The iterative computation model described above 
is implemented via the core computational method,
\textbf{\lstinline{compute}}.
To use the \textbf{\lstinline{compute}} method to orchestrate their iterative
computation,
users encode their vertex (resp., hyperedge) behavior in the form of a
\textbf{\lstinline{Program}} comprising a \textbf{\lstinline{Procedure}} for
consuming incoming messages, updating state, and producing outgoing messages,
as well as a \textbf{\lstinline{MessageCombiner}} for aggregating messages
destined to a common hyperedge (resp., vertex).
The \textbf{\lstinline{Context}} provides methods that enable the
\textbf{\lstinline{Procedure}} to update vertex (resp., hyperedge) state,
and to send messages to neighboring hyperedges (resp., vertices). 
When a vertex (resp., hyperedge) broadcases a message, the message is sent to all hyperedges (resp., vertices) 
to which the vertex (resp., hyperedge) is incident on.

In this model, hyperedges are elevated to first-class status; they can
maintain their state, carry out computation, and send messages just as
vertices do.
The \mesh{} API therefore meets our expressiveness requirements.
The generality and conciseness of the API aid in making the API easy to use.

To further improve ease of use, we observe that, in many cases, it is possible
to determine the \textbf{\lstinline{MessageCombiner}} automatically based on the
message types.
We implement this convenient feature
using Twitter's Algebird\footnote{\texttt{https://github.com/twitter/algebird}} 
library, and allow programmers to enable it with a single
\textbf{\lstinline{import}} directive.
With this feature enabled, users need only specify a \textbf{\lstinline{Procedure}}.

\subsection{Example \mesh{} Applications}
\label{sec:mesh-apps}

We next show how we can use the \mesh{} API to implement some of the algorithms discussed in Section~\ref{sec:algos}.
Listing~\ref{listing:pr} shows 
the implementation of a hypergraph variant of the
PageRank algorithm which computes ranks for both hyperedges and vertices iteratively.
As seen from the pseudocode, it is fairly simple to implement the algorithm, requiring only a few lines of code.
As shown in Listing~\ref{listing:prent},
a richer version of PageRank which also computes the entropy of each hyperedge (PageRank-Entropy, as described in Section~\ref{subsec:pr}) requires 
a simple three-line helper function to compute entropy and changes to only a few other lines (broadcast and rank computation). 

Implementing a Label Propagation Algorithm is also simple. 
Listing~\ref{listing:lp} shows how concisely we can implement
this algorithm using our API. Finally, Listing~\ref{listing:sp} shows an implementation of the Single Source Shortest Paths algorithm using our API. 
In this algorithm, the attribute values of both hyperedges and vertices are updated incrementally: if the path length increases, this update is broadcast to neighbors.
The algorithm terminates when the attribute values of every hyperedge and vertex are less than the updated values they received.
The major difference between the Shortest Path algorithm and the other algorithms above is that only a subset of hyperedges and vertices are active during any iteration (ones which were updated with newValue in the previous iteration). In contrast, every hyperedge and vertex is active in every iteration for the other algorithms. Note that for Label Propagation, PageRank and Shortest Paths, the
\textbf{\lstinline{MessageCombiner}} is derived automatically.

%---------- Approximate column width for IPDPS -----------|
\begin{lstlisting}[
style=myScalastyle,
caption={PageRank algorithm implementation.},
label={listing:pr},
emph={Procedure, become, broadcast},
emphstyle={\textbf}
]
	// Vertex procedure
	val vertex: Procedure =
	(superstep, id, attr, msg, ctx) => {
		val (totalWeight, rank) = msg
		val (vertexData, _) = attr
		// alpha = 0.15 (input from user)
		val newRank = alpha + (1.0 - alpha) * rank
		// Set its own vertex value
		ctx.become((vertexData, newRank))
		// Send data to neighbor hyperedges
		ctx.broadcast(newRank / totalWeight)
	}
	// Hyperedge procedure
	val hyperedge: Procedure =
	(superstep, id, attr, msg, ctx) => {
		val ((cardinality, weight), _) = attr
		val newRank = msg * weight
		// Set its own hyperedge value
		ctx.become(((card, weight), newRank))
		// Send data to neighbor vertices
		ctx.broadcast((weight, newRank / cardinality))
	}
\end{lstlisting}

%---------- Approximate column width for IPDPS -----------|
\begin{lstlisting}[
style=myScalastyle,
caption={PageRank-Entropy algorithm implementation: Changes from PageRank implementation are shown here.},
label={listing:prent},
emph={Procedure, become, broadcast},
emphstyle={\textbf}
]
	// Entropy function
	def entropy(ranks: Seq[Double]): Double = {
		val totalRank = ranks.sum
		val normalizedRanks = ranks.map(_ / totalRank)
		normalizedRanks.map { 
			p => p * math.log(1/p) }.sum / math.log(2)
		}
	// Vertex procedure
	val vertex: Procedure =
	(superstep, id, attr, msg, ctx) => {
	    ...
	    ctx.broadcast(Seq(newRank -> totalWeight))
	}
	// Hyperedge procedure
	val hyperedge: Procedure =
	(superstep, id, attr, msg, ctx) => {
	    ...
	    val newRank = msg.map { 
	        case (rank, totalWeight) =>
	        rank * weight / totalWeight 
	    }.sum
	    val newEnt = entropy(msg.map(_._1))
	    ...
	}
\end{lstlisting}

%---------- Approximate column width for IPDPS -----------|
\begin{lstlisting}[
style=myScalastyle,
caption={Label Propagation algorithm implementation.},
label={listing:lp},
emph={Procedure, become, broadcast},
emphstyle={\textbf}
]
	// Vertex procedure
	val vertex: Procedure =
	(superstep, id, attr, msg, ctx) => {
		val (vertexData, _) = attr
		val newLabel =
			if (superstep == 0) id
			else msg.max()
			// Set its own vertex value
			ctx.become((vertexData, newLabel))
			// Send data to neighbor hyperedges
			ctx.broadcast((newLabel))
	}
	// Hyperedge procedure
	val hyperedge: Procedure =
	(superstep, id, attr, msg, ctx) => {
		val (hyperedgeData, _) = attr
		val newLabel = msg.max()
		// Set its own hyperedge value
		ctx.become((hed, newLabel))
		// Send data to neighbor vertices
		ctx.broadcast((newLabel))
	}
\end{lstlisting}

%---------- Approximate column width for IPDPS -----------|
\begin{lstlisting}[
style=myScalastyle,
caption={Shortest Path algorithm implementation.},
label={listing:sp},
emph={Procedure, become, broadcast},
emphstyle={\textbf}
]
	// Vertex procedure
	val vertex: Procedure =
	(superstep, id, attr, msg, ctx) => {
	    val (vertexData, currentHop) = attr
	    val (_, newHop) = msg
	    if (currentHop > newHop) {
		    // Set its own vertex value
		    ctx.become((vertexData, newHop))
		    // Send data to neighbor hyperedges
		    ctx.broadcast(newHop + 1.0)
		}
	}
	// Hyperedge procedure
	val hyperedge: Procedure =
	(superstep, id, attr, msg, ctx) => {
	    val ((card, _), currentHop) = attr
	    val newHop = msg
	    if (currentHop > newHop) {
		    // Set its own hyperedge value
		    ctx.become((card, _), newHop)
		    // Send data to neighbor vertices
		    ctx.broadcast((_, newHop))
		}
	}

\end{lstlisting}
            % label: sec:api
\section{Implementation}
\label{sec:implementation}

In order to implement a scalable hypergraph processing system, we must address
two key challenges: how to represent the hypergraph, and how to partition this
representation for distributed computation.
As discussed in Section~\ref{sec:overview}, \mesh{} leverages the capabilities of 
an underlying graph processing system to address these challenges. 
Thus, it converts a hypergraph into an underlying graph representation,
and utilizes a graph partitioning framework to 
implement a variety of hypergraph partitioning algorithms.
We next discuss the design choices  and the tradeoffs in making these decisions,
as well as our implementation on top of GraphX.

\subsection{Representation}
The first question we must address is how to represent a hypergraph as an underlying graph that is understandable by a graph processing system, and we consider two alternatives here.

%\paragraph
\subsubsection{Clique-Expanded Graph}
\begin{figure}%[htbp]
	\centering
	\subfigure[Clique-expanded.]{
		\raisebox{1.5em}{
			\includegraphics[width=0.40\columnwidth]{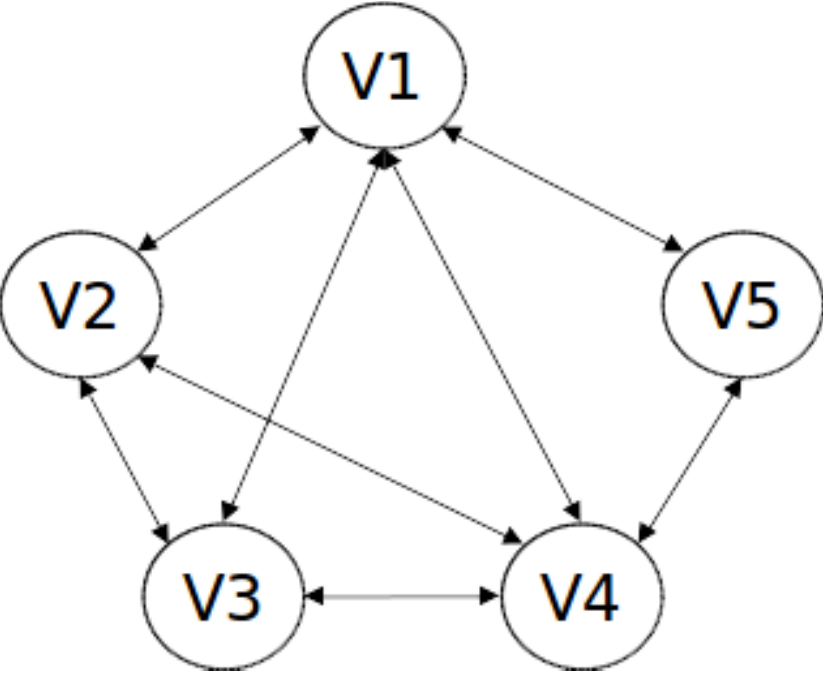}
		}
		\label{fig:clique-expanded-representation}
	}
	\hspace{3em}
	\subfigure[Bipartite.]{
		\includegraphics[height=1.5in]{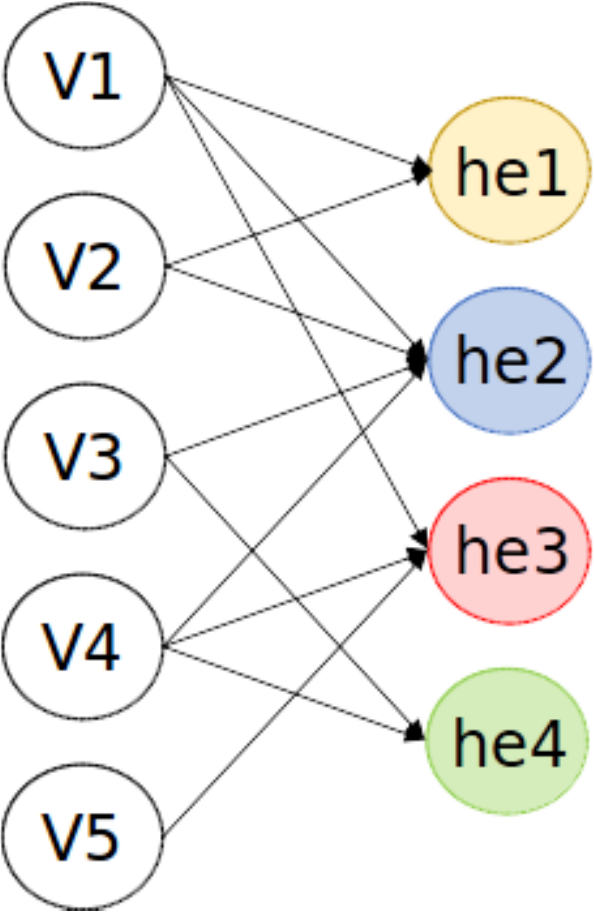}
		\label{fig:bipartite-representation}
	}
	\vspace{-0.25em}
	\caption{Underlying graph representations of the hypergraph in
		Figure~\ref{fig:three_authors_hypergraph}}
	\vspace{-1em}
	\label{fig:underlying-representations}
\end{figure}

One possibility is
to represent a hypergraph as a simple
graph by expanding each hyperedge into a clique of its members.
We refer to this representation as the \emph{clique-expanded graph}.
Figure~\ref{fig:clique-expanded-representation} shows the clique-expanded representation for the example hypergraph shown in Figure~\ref{fig:three_authors_hypergraph}.
In order to enable this representation, our \textbf{\lstinline{HyperGraph}}
interface provides a \textbf{\lstinline{toGraph}} transformation method, which
logically replaces the connectivity structure of the hypergraph with edges
rather than hyperedges.
The attributes of an edge from $v_1$ to $v_2$ are determined by user-defined
functions applied to the set of all hyperedges common to $v_1$
and $v_2$.
Applying this transformation to produce a clique-expanded graph may be
costly---even prohibitively so---in terms of both space and time.

Another major disadvantage of the clique-expanded graph is its limited
applicability.
Because hyperedges do not appear in this representation, it is only
appropriate for algorithms that do not modify hyperedge state,
and thus, for instance cannot be used for our Label Propagation algorithm.
Further, the hyperedge and vertex programs must meet additional requirements,
such as sending the same message type in both directions.
Overall, therefore, this representation is best viewed as a potential
optimization for a small set of use cases rather than a general approach.

%\paragraph
\subsubsection{Bipartite Graph}
An alternative approach is to represent the hypergraph internally as a {\em bipartite
graph}, where one partition comprises exclusively vertices, and the other
exclusively hyperedges, with low-level graph edges connecting hyperedges to
their constituent vertices.
Figure~\ref{fig:bipartite-representation} shows the bipartite representation for the example hypergraph from Figure~\ref{fig:three_authors_hypergraph}.
This representation can concisely encode any hypergraph, and it allows us to
run programs that treat both vertices and hyperedges as first-class
computational entities.
By using directed edges (in our implementation exclusively from vertices to
hyperedges), we provide a means to differentiate between vertices and hyperedges.
Due to the general expressive power of this representation, {\em we focus our
attention throughout this paper on its efficient implementation in a graph
processing system.}

\subsection{Partitioning}
\subsubsection{Challenges}
To scale to large hypergraphs, it is essential to distribute computation across
multiple nodes.
The decision of how to partition the underlying representation can
significantly affect performance, in terms of both computational load and
network I/O.
An effective partitioning algorithm---whether for a graph or a hypergraph---must
simultaneously balance computational load and minimize communication.
Hypergraph partitioning, however, presents several challenges beyond those for 
partitioning graphs.

For one, hypergraphs contain two distinct sets of entities: vertices and
hyperedges.
In general, these two sets can differ significantly in terms of their size,
skew in cardinality/degree%
\footnote{The \emph{degree} of a vertex denotes the number of hyperedges of which that
vertex is a member.
Similarly, the \emph{cardinality} of a hyperedge denotes the number of
vertices belonging to that hyperedge.
}%
, and associated computation.
Further, \mesh{} computation runs on only one of these sets at a time.
An effective partitioning algorithm must therefore
\emph{differentiate between hyperedges and vertices}.

At the same time, hyperedge and vertex partitioning are fundamentally
interrelated; an effective algorithm must \emph{holistically partition hyperedges
and vertices}.
For example, an algorithm that partitions hyperedges without regard to vertex
partitioning may achieve good computational load balance,
but will suffer from excessive network I/O.

\subsubsection{Algorithms}

\mesh{} utilizes the underlying graph partitioning framework to implement 
hypergraph partitioning algorithms. 
Many graph processing frameworks either partition vertices (cutting edges%
\footnote{Note that we use ``edge'' to refer to an edge in the {\em underlying} graph representation, and it is not to be confused with a hyperedge in the provided hypergraph.}%
) or
partition edges (thus cutting vertices) across machines.
Many current systems~\cite{gonzalez2014,gonzalez2012} use edge partitioning since
it has been shown to be more efficient for many real-world graphs. 
In what follows, we describe mapping hypergraph partitioning algorithms to
such edge partitioning graph algorithms. 
We expect that mapping to vertex partitioning algorithms could be done in a similar
fashion, and we leave such mapping as future work.

Concretely, we assume the underlying graph partitioning framework partitions 
the set of edges, while replicating
each vertex to every partition that contains edges incident on that
vertex.
In our bipartite graph representation, edges are directed exclusively from
(hypergraph) vertices to hyperedges.
As a result, if we partition
based only on the source (resp., destination) of an edge, hypergraph vertices
(resp., hyperedges) are each assigned to a unique partition, while hyperedges
(resp., vertices) will be replicated---i.e., ``cut''---across several
partitions.
If we choose the partition for an edge
based on both its source and destination, then both vertices and hyperedges are
effectively cut.

Any graph partitioning algorithm leads to a tradeoff between balancing
computational load and minimizing network communication.
While balancing the number of edges across machines could lead to good load balance, 
a high degree of replication of vertices can lead to
increased network I/O and execution time due to increased syncing and state updates.
In order to distribute a hypergraph, however, replication is unavoidable.
The goal is therefore to choose which set(s) (vertices or hyperedges) to
cut, and how to partition the other set so as to balance computational load
while minimizing replication.

\begin{figure}%[htbp]
	\centering
	\subfigure[Random Vertex-cut]{
		\includegraphics[height=1.5in]{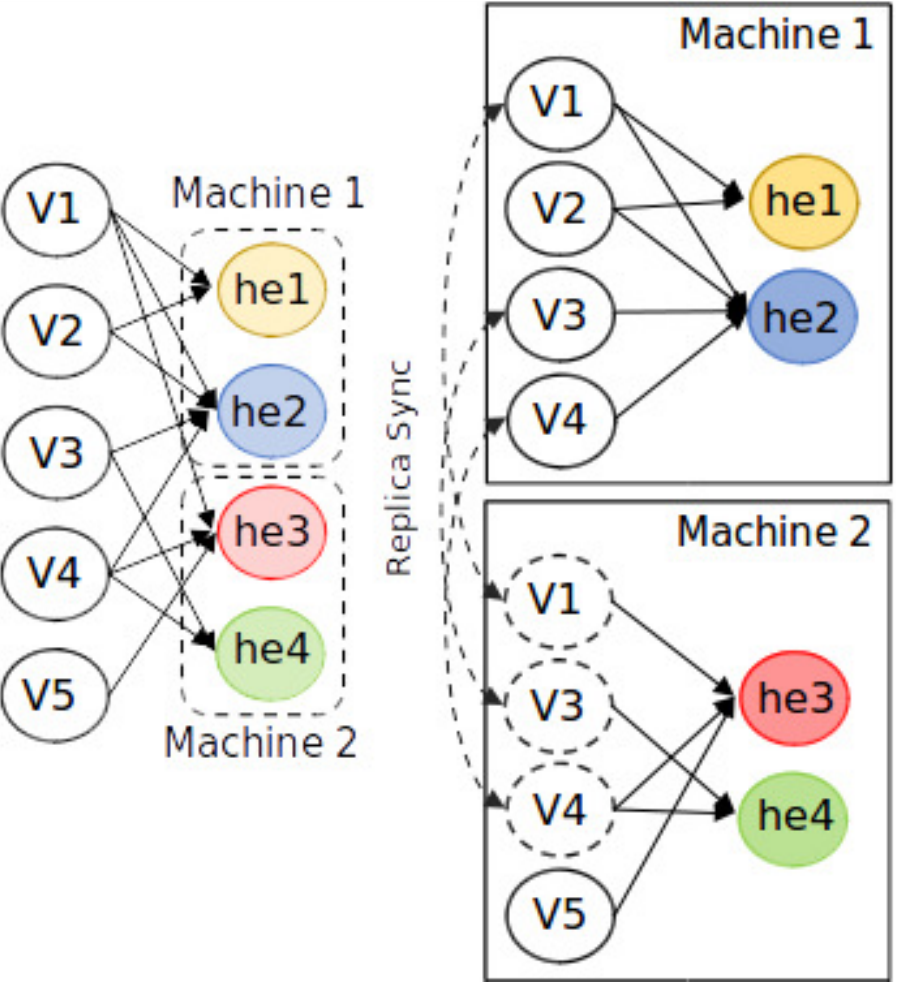}
		\label{fig:randomV}
	}
	\subfigure[Random Hyperedge-cut]{
		\raisebox{0.5em}{
			\includegraphics[width=0.45\columnwidth]{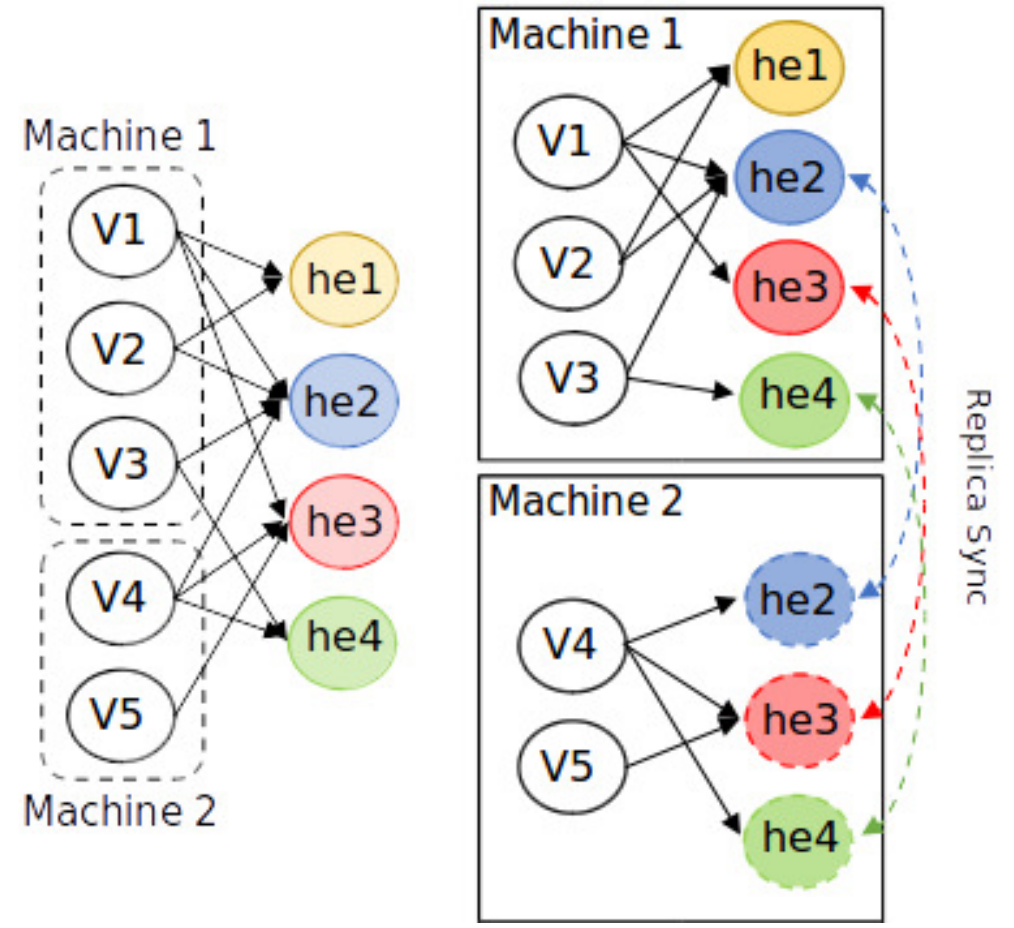}
		}
		\label{fig:randomHE}
	}
	\vspace{-0.25em}
	\caption{Random partitioning strategies.}
	\vspace{-1em}
	\label{fig:random_strategies}
\end{figure}

\begin{figure}%[htbp]
	\centering
	\subfigure[Hybrid Vertex-cut (cutoff = 3)]{
		\includegraphics[height=1.5in]{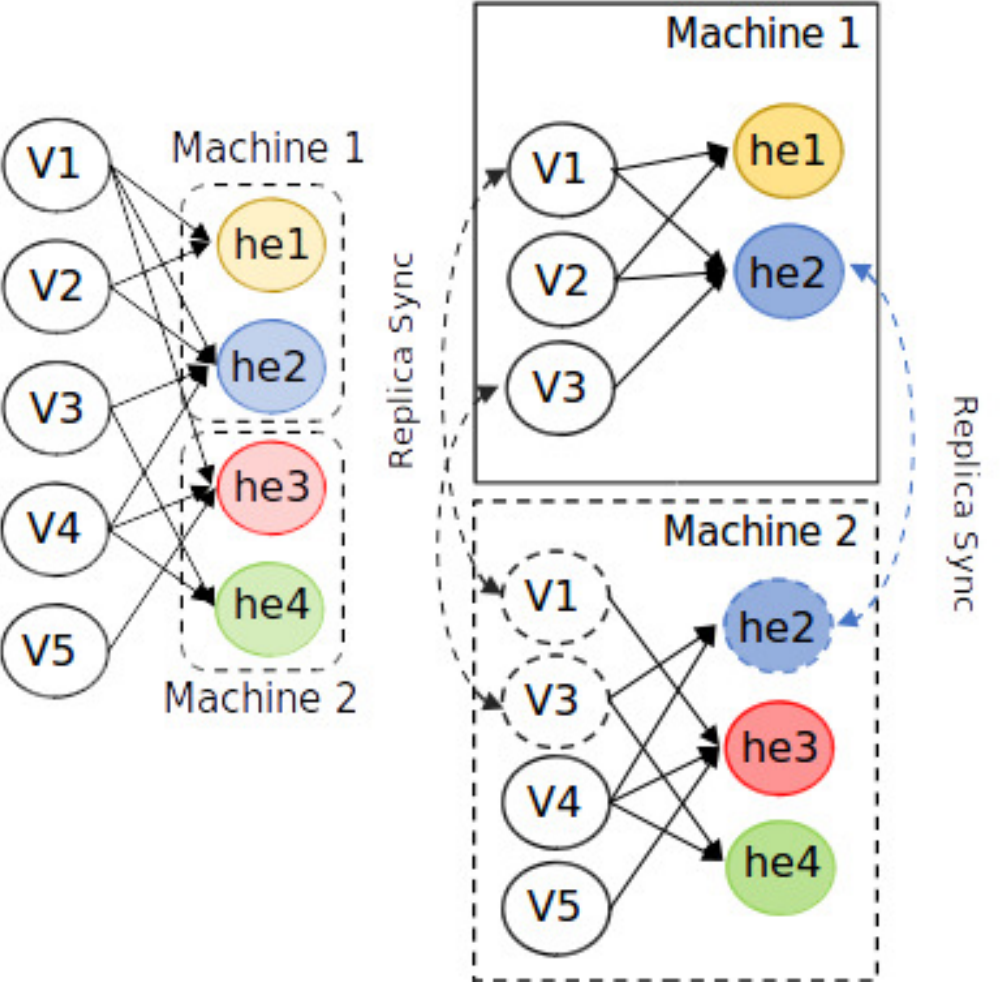}
		\label{fig:hybridV}
	}
	\subfigure[Hybrid Hyperedge-cut (cutoff = 2)]{
		\raisebox{0.5em}{
			\includegraphics[width=0.45\columnwidth]{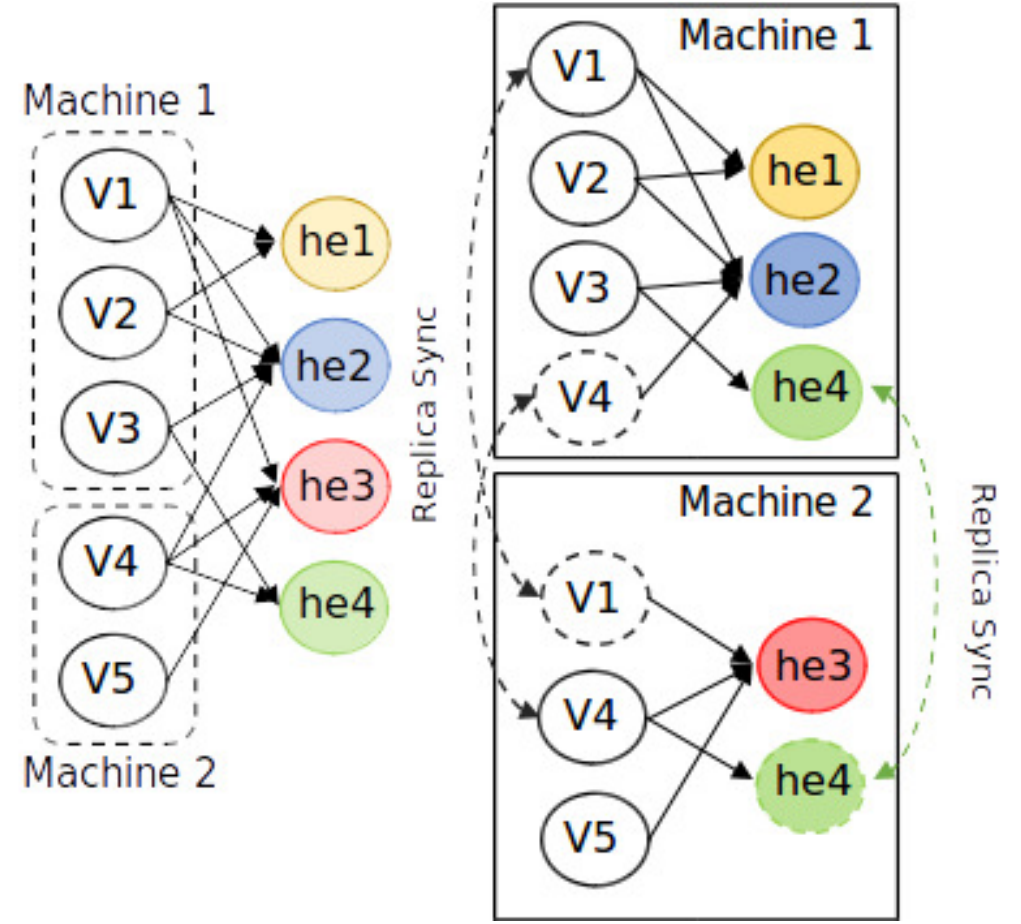}
		}
		\label{fig:hybridHE}
	}
	\vspace{-0.25em}
	\caption{Hybrid partitioning strategies.}
	\vspace{-1em}
	\label{fig:hybrid_strategies}
\end{figure}

\begin{figure}%[htbp]
	\centering
	\subfigure[Greedy Vertex-cut]{
		\includegraphics[width=0.75\columnwidth]{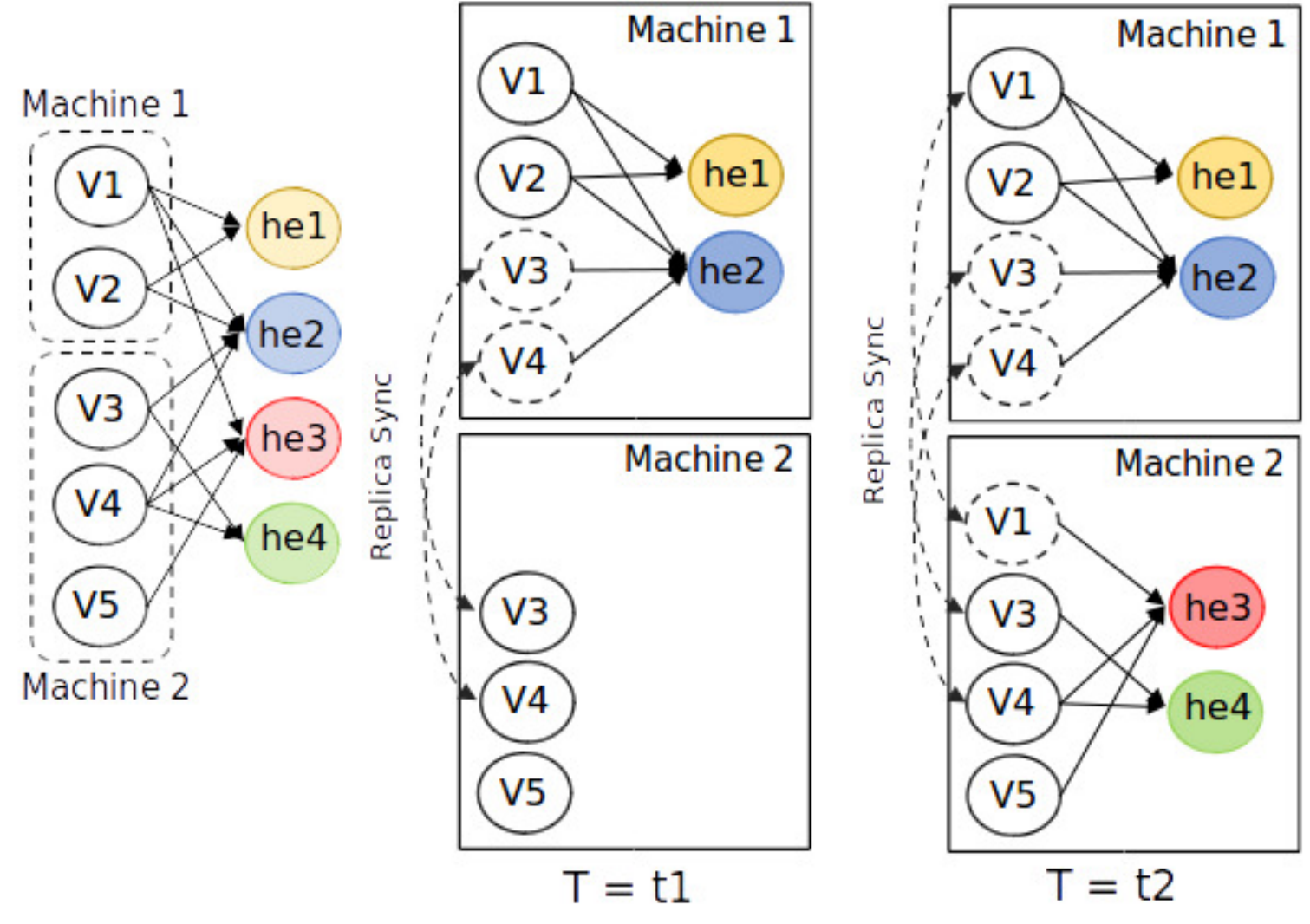}
		\label{fig:greedyV}
	}
	
	\subfigure[Greedy Hyperedge-cut]{
		\hspace{2.5em}
		\includegraphics[width=0.80\columnwidth]{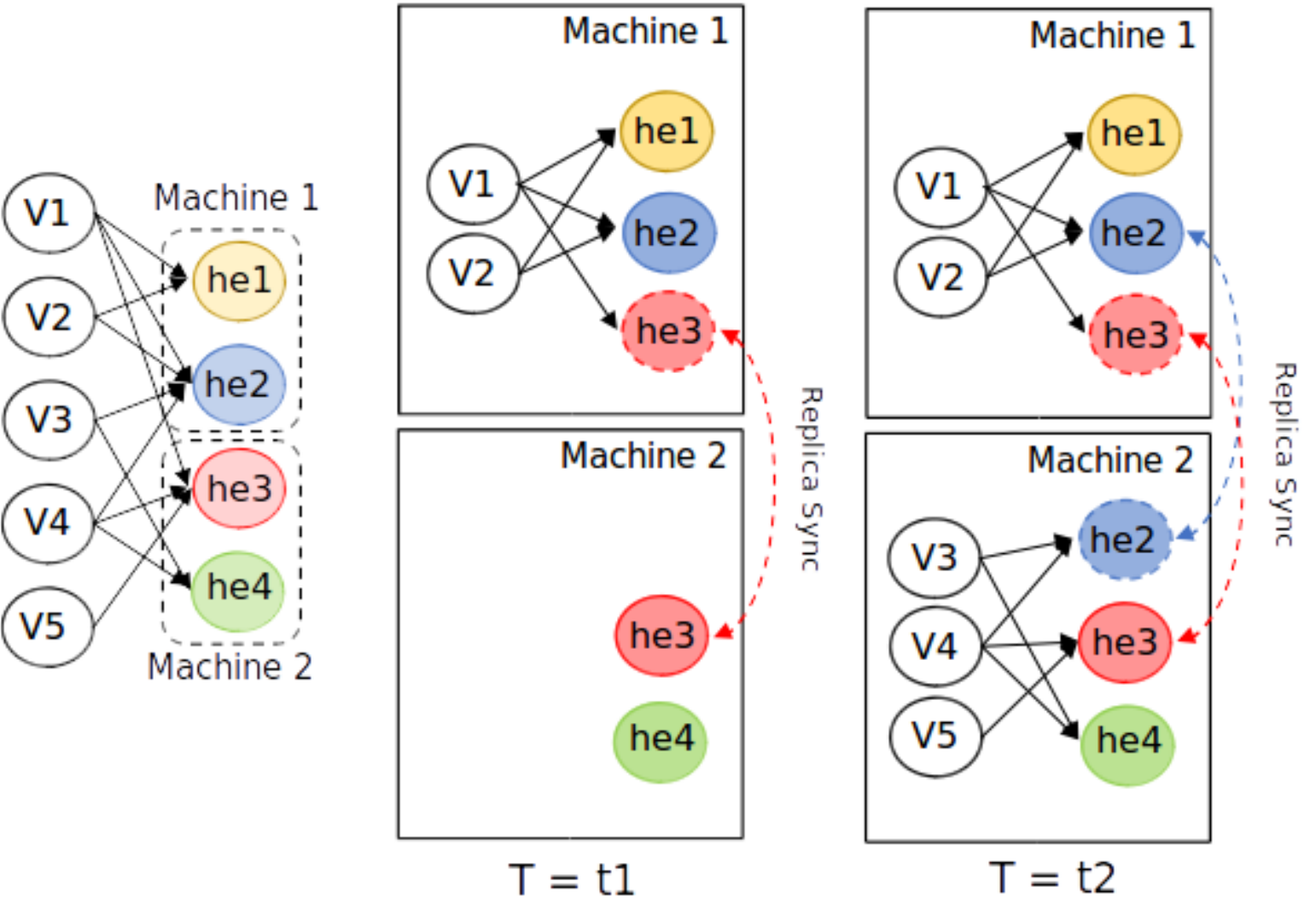}
		\label{fig:greedyHE}
	}
	\vspace{-0.25em}
	\caption{Greedy partitioning strategies.}
	\vspace{-1em}
	\label{fig:greedy_strategies}
\end{figure}

We explore a range of alternative partitioning algorithms that approach this
goal from different angles.
These algorithms fall into three classes: \emph{Random},  \emph{Greedy}, and \emph{Hybrid}. 
We illustrate each of these algorithms by showing how they would partition our example hypergraph bipartite representation (Figure~\ref{fig:bipartite-representation}) on two machines.

\paragraph{Random}
We explore three Random partitioning algorithms.
The {\em Random Vertex-cut} algorithm hash-partitions bipartite graph edges
based on their destination (i.e., by hyperedge), effectively cutting
hypergraph vertices. 
For example, in Figure~\ref{fig:randomV}, the algorithm assigns each hyperedge to either machine1 or machine2 using a hash function. It then assigns a replica of each vertex to every machine which contains its 
incident hyperedges. E.g.: $v_1$ is assigned to machines 1 and 2, as it is incident on $he_1$ and $he_2$ (on machine 1), and $he_3$ (on machine 2).    
The {\em Random Hyperedge-cut} algorithm, on the other hand, partitions
hyperedges and cuts vertices, as shown in Figure~\ref{fig:randomHE}. 

The {\em Random Both-cut} algorithm hash-partitions bipartite graph edges
by both their source and destination, effectively cutting both
vertices and hyperedges.

\paragraph{Hybrid}
The Hybrid algorithms we consider are based on the balanced $\rho$-way
hybrid cut from PowerLyra~\cite{chen2015}.
These algorithms cut both vertices and hyperedges, but unlike Random Both-cut,
they differentiate between vertices and hyperedges in doing so.
The {\em Hybrid Vertex-cut} variant cuts vertices while partitioning hyperedges, except that it also cuts hyperedges with high cardinality (greater than 100 in our experiments). 
In Figure~\ref{fig:hybridV}, the algorithm cuts vertices $v_1$ and $v_3$ while partitioning hyperedges and cuts hyperedge $he_2$ since it has cardinality greater than the cutoff value of 3 in the example. 
Similarly, the {\em Hybrid Hyperedge-cut} variant cuts hyperedges while also cutting
high-degree vertices, as shown in Figure~\ref{fig:hybridHE}.

\paragraph{Greedy}
Based on the Aweto~\cite{chen2014} algorithm, the Greedy algorithms that we
consider holistically partition hypergraphs with the goal of reducing
replication and the resulting synchronization overhead.
At a high level, the {\em Greedy Vertex-cut} variant aims to assign each hyperedge
to a lightly-loaded partition with a large ``overlap'' between the vertices in
that hyperedge and the vertices with replicas already on that partition based
on (a heuristic estimate of) the assignments already made.
(For a more rigorous definition, see~\cite{chen2014}.)

Figure~\ref{fig:greedy_strategies} illustrates the details of Greedy partitioning strategies,
showing how the strategies incrementally assign vertices and hyperedges. % over two different time steps.
In Figure~\ref{fig:greedyV}, the Greedy Vertex-cut algorithm first hash-partitions the bipartite graph edges based on their vertices and then assigns one hyperedge at a time to an appropriate machine. T = t1 in Figure~\ref{fig:greedyV} shows the intermediate state after hyperedges $he_1$ and $he_2$ have been assigned. Hyperedge $he_1$ is assigned to machine 1 because the machine contains maximum number of incident vertices at this time. Because load and overlap are even across machines at this time, hyperedge $he_2$ is randomly assigned to machine 1, and $v_3$ and $v_4$ are cut accordingly. T = t2 in the figure shows the final state of the partitioning. Hyperedges $he_3$ and $he_4$ are assigned to machine 2 because it contains 
maximum overlapping edges and the machine is also lightly loaded at this time.
The {\em Greedy Hyperedge-cut} variant in Figure~\ref{fig:greedyHE}
works similarly, except 
it assigns vertices based on
the overlap between their incident hyperedges and the hyperedges already
assigned to each partition.

\subsection{Implementation on GraphX}

As mentioned in Section~\ref{sec:overview}, 
we have implemented a \mesh{} prototype on top of the GraphX~\cite{gonzalez2014} graph processing system.
GraphX provides a graph representation consisting of a VertexRDD and an EdgeRDD which internally extend Spark's Resilient Distributed Datasets (RDDs), an immutable and partitioned collection of elements~\cite{zaharia2012}. VertexRDD contains information about vertex ids and vertex attributes. EdgeRDD contains information about edges (pairs of vertices) and edge attribute properties. % for the edge between them.

In our implementation, Hypergraph contains an additional HyperEdgeRDD which is similar to VertexRDD and contains information about hyperedge ids and hyperedge attributes. Moreover, EdgeRDD now contains information about (vertex, hyperedge) pairs and the attribute properties for the relation between them.  
We can represent a hypergraph using a clique representation by mapping each hyperedge to a clique of its incident vertices. Similarly, we can represent a hypergraph as a bipartite graph by creating edges between vertices and their hyperedges. 

GraphX uses an edge-partitioning algorithm for partitioning the graph across machines.
In GraphX, the \textbf{\lstinline{PartitionStrategy}} considers each edge in
isolation, as in Listing~\ref{listing:graphx-old}.

\begin{lstlisting}[
style=myScalastyle,
caption={Original GraphX partitioning abstraction.},
label={listing:graphx-old},
basicstyle={\footnotesize\ttfamily}
]
def getPartition(
	src: VertexId,
	dst: VertexId,
	numPart: PartitionId): PartitionId
\end{lstlisting}

\begin{lstlisting}[
style=myScalastyle,
caption={Extended GraphX partitioning abstraction.},
label={listing:graphx-new},
basicstyle={\footnotesize\ttfamily}
]
def getAllPartitions[VD, ED](
graph: Graph[VD, ED],
numPartitions: PartitionId,
degreeCutoff: Int)
: RDD[((VertexId, VertexId), PartitionId)]
\end{lstlisting}

We use the built-in GraphX partitioning algorithms to implement the baseline Random partitioning algorithms described above. 
Our Greedy and Hybrid algorithms, however, require a broader view of the graph
(to compute ``overlap'' and degree/cardinality, respectively).
To satisfy this requirement, we extend the
\textbf{\lstinline{PartitionStrategy}} by adding a new
\textbf{\lstinline{getAllPartitions}} method that allows partitioning
decisions to be made with awareness of the full graph, as shown in
Listing~\ref{listing:graphx-new}.

Unlike the getPartition method of GraphX, which receives source and destination vertices for an edge and returns partition number for that edge, the getAllPartitions method receives property graph corresponding to a pair of VertexRDD and EdgeRDD and returns an RDD which maps source and destination vertices with their associated partition number as shown in Listing~\ref{listing:graphx-new}.
Next, we discuss how Hybrid and Greedy partitioning algorithms use this extended partitioning interface.

%---------- Approximate column width for IPDPS -----------|
\begin{lstlisting}[
style=myScalastyle,
caption={Hybrid Vertex-cut PartitionStrategy.},
label={listing:hybrid-hypervertex-cut}
]
// mPrime: large prime number for better random assignment
val in_degrees = graph.edges.map(
	e => (e.dstId, (e.srcId, e.attr))).
	join(graph.inDegrees.map(e => (e._1, e._2))
)
in_degrees.map { e =>
	var part: PartitionID = 0
	if (Degree > degreeCutoff) {
		part = ((math.abs(srcId) * mPrime) % numParts).toInt
	} else {
		part = ((math.abs(dstId) * mPrime) % numParts).toInt}
	((srcId, dstId),part))
}
\end{lstlisting}

%---------- Approximate column width for IPDPS -----------|
\begin{lstlisting}[
style=myScalastyle,
caption={Greedy Vertex-cut PartitionStrategy.},
label={listing:greedy-hypervertex-cut}
]
// Count the number of edges corresponding to dstId
val groupedDst = graph.edges.map{
	e => (e.dstId, e.srcId)}.groupByKey
// Count the overlap for dstId with each partition.
val dstEdgeCount = groupedDst.map{ e =>
	var dstOverlap = new Array[Long](numParts)
	e._2.map{srcs => dstOverlap(
		(math.abs(srcs * mPrime) % numParts).toInt) += 1}
	(e._1, dstOverlap)
}
// Iterate to find most overlapping partitions
val FinalDstAssignment = dstEdgeCount.map { e =>
	var mostOverlap : Double = 
	    src.apply(part) - math.sqrt(1.0*current_load(part))
	for (cur <- 0 to numParts-1){
	    val overlap : Double = 
		    src.apply(cur) - math.sqrt(1.0*current_load(cur))
	    if (overlap > mostOverlap){
		    part = cur
		    mostOverlap = overlap
	    }
	}
	(dst, part)
}
\end{lstlisting}

Hybrid Vertex-cut PartitionStrategy uses different partitioning policy for low and high degree vertices. In this PartitionStrategy, if the cardinality of a particular hyperedge exceeds the provided threshold (degreeCutoff), it cuts the hyperedge and partitions it based on the hashing of source vertex; otherwise, it partitions based on the hashing of destination vertex as shown in Listing~\ref{listing:hybrid-hypervertex-cut}. It is done to reduce communication overhead due to high degree hyperedges.

Listing~\ref{listing:greedy-hypervertex-cut} shows Greedy Vertex-cut PartitionStrategy which uses overlap and load for partitioning. This PartitionStrategy considers one hyperedge at a time and greedily chooses a partition that the hypervertices contained in this hyperedge most overlapped with. Additionally, if the load on the chosen partition is high, it picks the next most overlapped partition. 
 % label: sec:implementation
\section{Evaluation}
\label{sec:evaluation}

\begin{table*}[htbp]
  \centering
  \small
  \caption{Datasets used in our experiments.\label{tab:datasets}}
  \begin{tabular}{ | l | c | c | c | c | c | c | }
    \hline
    Dataset & \# Vertices & \# Hyperedges & Max. Degree & Max. Cardinality & \# Bipartite Edges & \# Clique-Expanded Edges \\ \hline
    Apache & 3316 & 78,080 & 4,507 & 179 & 408,231 & 196,452 \\ \hline
    dblp & 899,393 & 782,659 & 368 & 2,803 & 2,624,912 & 21,707,067 \\ \hline
    Friendster & 7,944,949 & 1,620,991 & 1,700 & 9,299 & 23,479,217 & \textit{10.3 billion (approximate)} \\ \hline
    Orkut & 2,322,299 & 15,301,901 & 2,958 & 9,120 & 107,080,530 & \textit{54.5 billion (approximate)} \\ \hline
  \end{tabular}
\end{table*}

\subsection{Experimental Setup}
\subsubsection{Deployment}
We implement and run our \mesh{} prototype on top of Apache Spark 1.6.0.
Experiments are conducted on a cluster of eight nodes, each with two Intel Xeon
E5-2620 v3 processors with 6 physical cores and hyperthreading enabled.
Each node has 64 GB physical RAM, and a 1 TB hard drive with at least 75\% free
space, and nodes are connected via gigabit ethernet. 
Input data are stored in HDFS 2.7.2, which runs across these same eight nodes.
We show both partitioning time and execution time in our results.

\subsubsection{Datasets}
As inputs for our experiments, we use publicly available data~\cite{snapnets} to build the
hypergraphs described in Table~\ref{tab:datasets}.
These datasets differ in their characteristics, such as size, relative number
of vertices and hyperedges, vertex degree/hyperedge cardinality distribution,
etc.

The Apache hypergraph, derived from the Apache Software Foundation subversion~\cite{apache-ssf}
logs, models collaboration on open-source software projects.
Each vertex represents a committer, and each hyperedge
represents a set of committers that have collaborated on one or more
files.

The dblp dataset describes more than one million publications, from which we
use authorship information to build a hypergraph model where vertices represent
authors and hyperedges represent collaborations between authors.

In the Friendster and Orkut hypergraphs, vertices represent individual users, and
hyperedges represent user-defined communities in the Friendster and Orkut social
networking sites, respectively.
Because membership in these communities does not require the same commitment as
collaborating on software or academic research, these hypergraphs have very
different characteristics from dblp and Apache, in particular in terms of the overall size of the data, and vertex degree and hyperedge cardinality. 
One difference between the two is that Friendster has many more vertices than
hyperedges, whereas the opposite is true for Orkut.

\subsubsection{Applications}
We use the four applications described in Section~\ref{sec:mesh-apps}
in our experiments: the Label Propagation
algorithm (Listing~\ref{listing:lp}), 
the two PageRank variants: PageRank (Listing~\ref{listing:pr}) and PageRank-Entropy (Listing~\ref{listing:prent}), and
the Shortest Paths algorithm (Listing~\ref{listing:sp}).

\subsection{Representation}
\begin{figure}
  \centering
  \includegraphics[width=0.85\columnwidth]{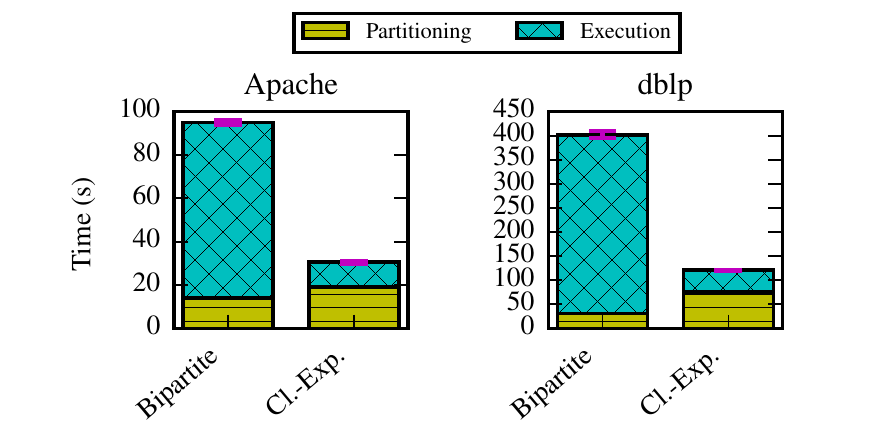}
  \caption{Run time comparison for bipartite and clique-expanded representations
    for the PageRank algorithms on the Apache and dblp datasets. 
Friendster and Orkut results are not shown since their clique-expansions could not be materialized.
  \label{fig:representation}}
  \vspace{-0.2cm}
\end{figure}

We first briefly explore the relative strengths and weaknesses of two main
representation alternatives: the clique-expanded graph and the bipartite graph.
Table~\ref{tab:datasets} shows the number of edges required for each of these
representations, while Figure~\ref{fig:representation} shows both partitioning
time and subsequent execution time for the PageRank algorithm for the Apache
and dblp hypergraphs.
Throughout this paper, we run each experiment multiple times and plot the mean,
with error bars denoting 95\% confidence intervals.

For the Apache hypergraph, the clique-expanded graph shows some promise in
terms of both space and time.
The clique-expanded representation uses only about 48\% as many edges as the
bipartite alternative, and although the initial partitioning phase (which
includes running the \textbf{\lstinline{toGraph}} transformation) is more
time-consuming for this representation, the execution is significantly faster.
For the dblp hypergraph, the clique-expanded representation again shows an
execution time advantage, but this comes at the cost of space overhead, as
this representation requires roughly 8x as many edges as the bipartite
alternative.

The clique-expansion can be thought of as a constant-folding optimization
applied at the time of constructing the representation.
Although this can be helpful in terms of execution time in some cases, its space overhead
can be large or even prohibitive.
In fact, for the Friendster and Orkut hypergraphs, we are unable to {\em even materialize} the
clique-expanded graphs on our cluster due to space limitations. 
As seen from Table~\ref{tab:datasets}, these datasets would result in approx. 10 billion and 54 billion clique-expanded edges respectively, which is orders of magnitude higher than that for their bipartite graph representation.

In addition to these scalability concerns, it is important to keep in mind
that the clique-expanded representation does not apply for all algorithms,
as discussed in Section~\ref{sec:implementation}.
For example, we cannot use this representation for our Label Propagation or
PageRank-Entropy algorithms, as these algorithms need explicit access to
hyperedge attributes.
Thus, while the clique-expansion representation might be beneficial for some
use cases (specific algorithms and small datasets), it is neither expressive enough
nor scalable in general.
Given these limitations, we focus on the more general alternative of the
bipartite representation throughout the remainder of our experiments.

\subsection{Partitioning}

\begin{figure*}[htpb]
  \centering
  \subfigure[dblp]{
    \includegraphics[width=0.60\columnwidth]{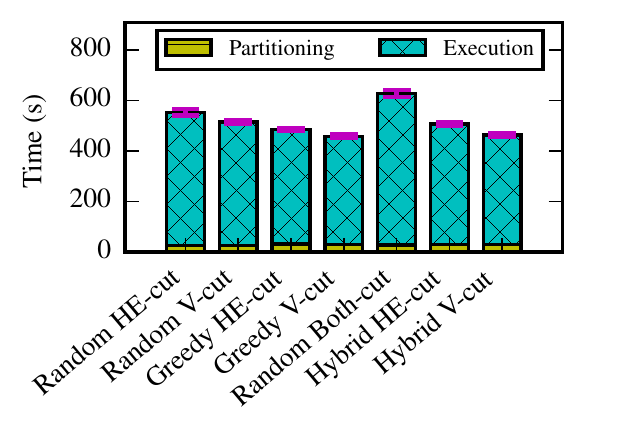}
    \label{fig:partitioning-lp-dblp}
  }
  \subfigure[Friendster]{
    \includegraphics[width=0.60\columnwidth]{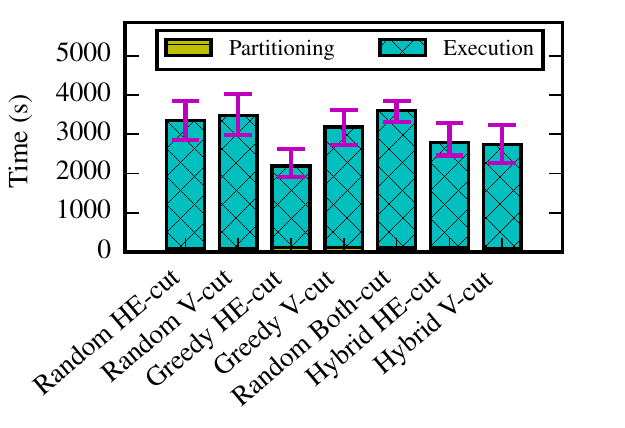}
    \label{fig:partitioning-lp-friendster}
  }
  \subfigure[Orkut]{
    \includegraphics[width=0.60\columnwidth]{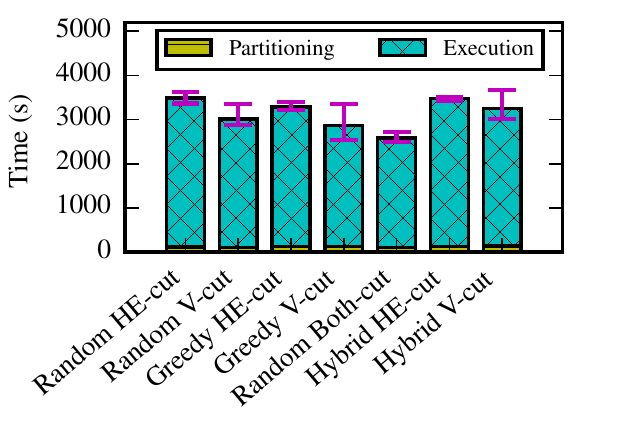}
    \label{fig:partitioning-lp-orkut}
  }
  \caption{Label Propagation: Partitioning and execution time using several
    partitioning algorithms in \mesh{}.\label{fig:partitioning-lp}}
\end{figure*}

% pr
\begin{figure*}[htpb]
  \centering
  \subfigure[dblp]{
    \includegraphics[width=0.60\columnwidth]{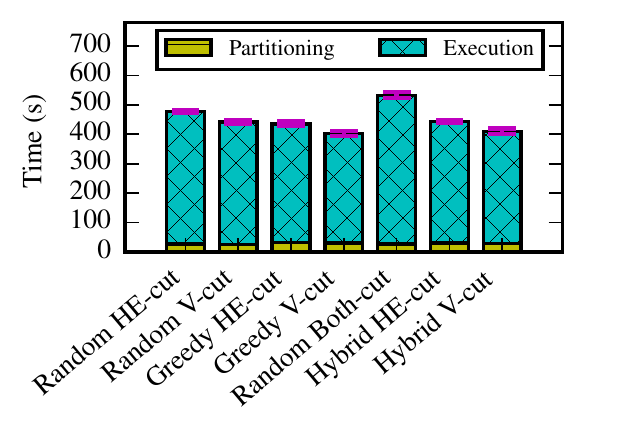}
    \label{fig:partitioning-pr-dblp}
  }
  \subfigure[Friendster]{
    \includegraphics[width=0.60\columnwidth]{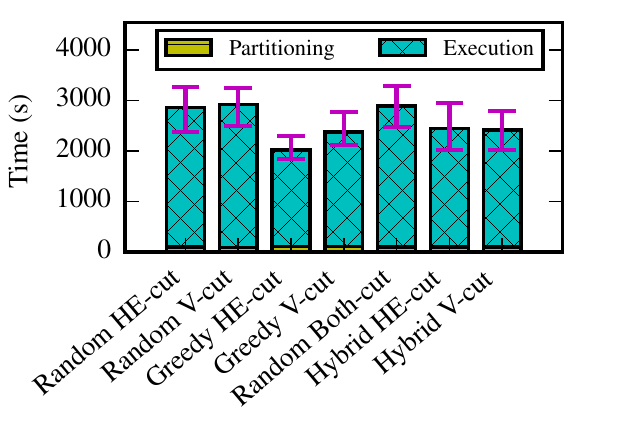}
    \label{fig:partitioning-pr-friendster}
  }
  \subfigure[Orkut]{
    \includegraphics[width=0.60\columnwidth]{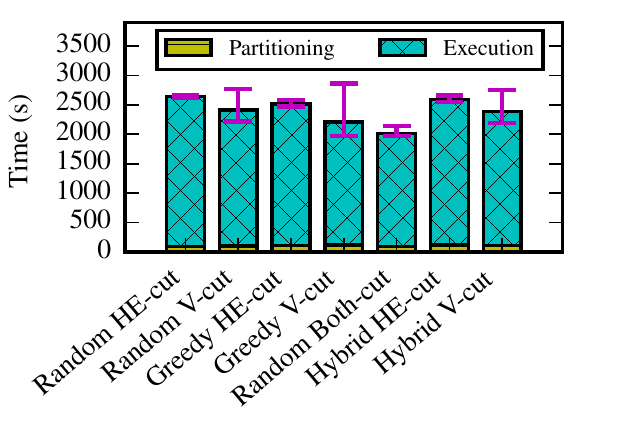}
    \label{fig:partitioning-pr-orkut}
  }
  \caption{PageRank: Partitioning and execution time using several
    partitioning algorithms in \mesh{}.\label{fig:partitioning-pr}}
\end{figure*}

% prEntropy
\begin{figure*}[htpb]
\centering
\subfigure[dblp]{
\includegraphics[width=0.60\columnwidth]{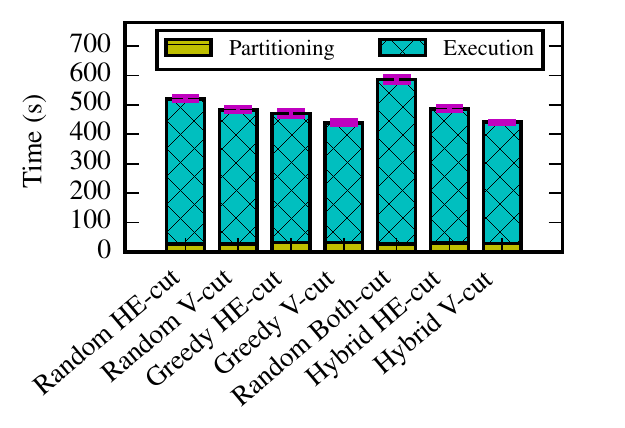}
\label{fig:partitioning-prEntropy-dblp}
}
\subfigure[Friendster]{
\includegraphics[width=0.60\columnwidth]{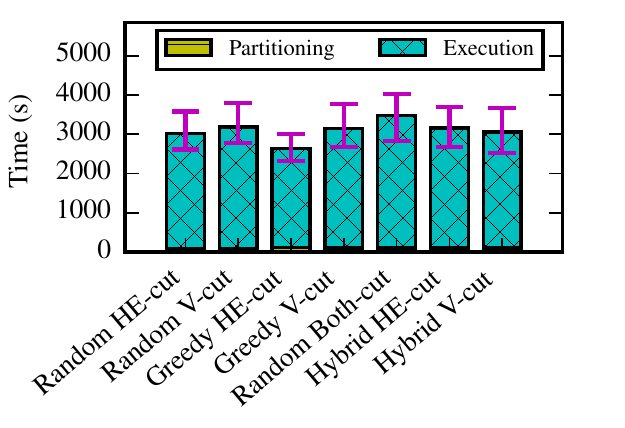}
\label{fig:partitioning-prEntropy-friendster}
}
\subfigure[Orkut]{
	\includegraphics[width=0.60\columnwidth]{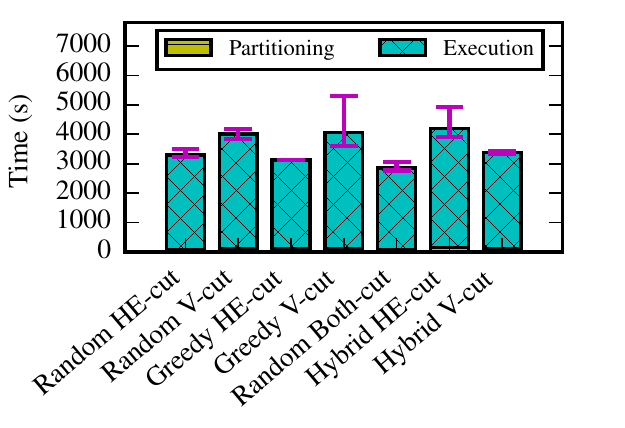}
	\label{fig:partitioning-prEntropy-orkut}
}
\caption{PageRank-Entropy: Partitioning and execution time using several
partitioning algorithms in \mesh{}.\label{fig:partitioning-prEntropy}}
\end{figure*}

% sp
\begin{figure*}[htpb]
	\centering
	\subfigure[dblp]{
		\includegraphics[width=0.60\columnwidth]{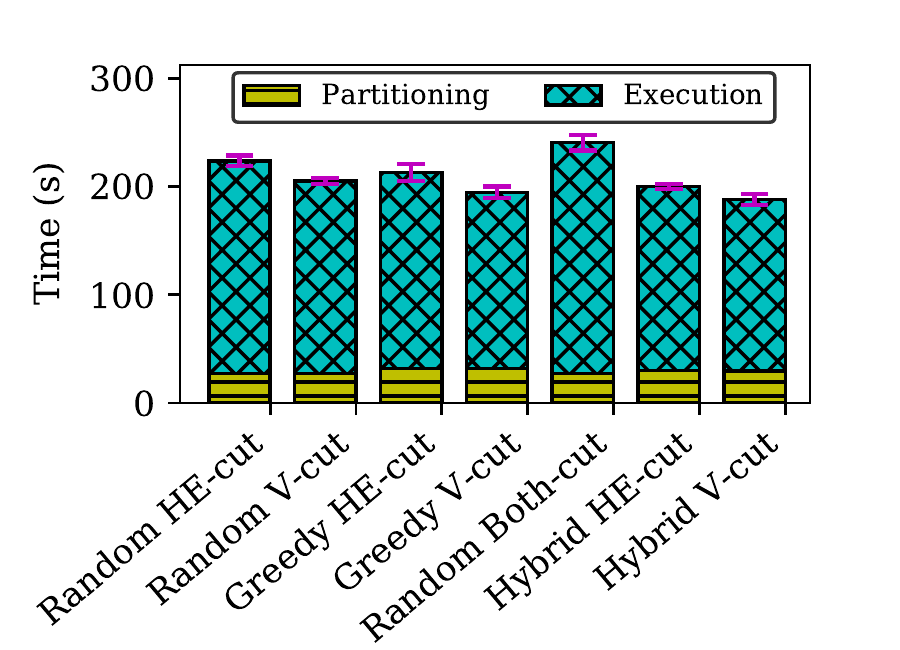}
		\label{fig:partitioning-sp-dblp}
	}
	\subfigure[Friendster]{
		\includegraphics[width=0.60\columnwidth]{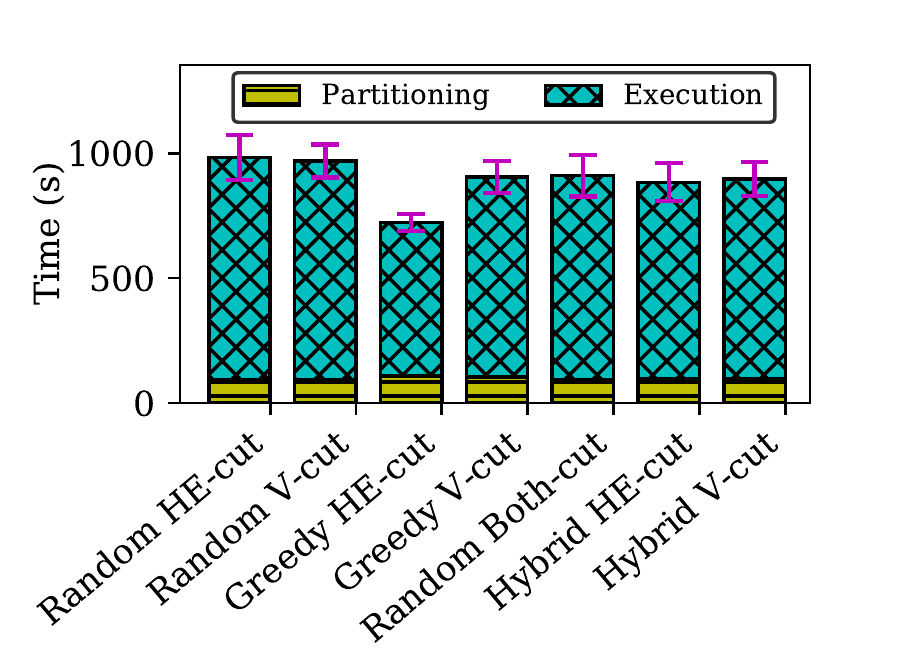}
		\label{fig:partitioning-sp-friendster}
	}
	\subfigure[Orkut]{
		\includegraphics[width=0.60\columnwidth]{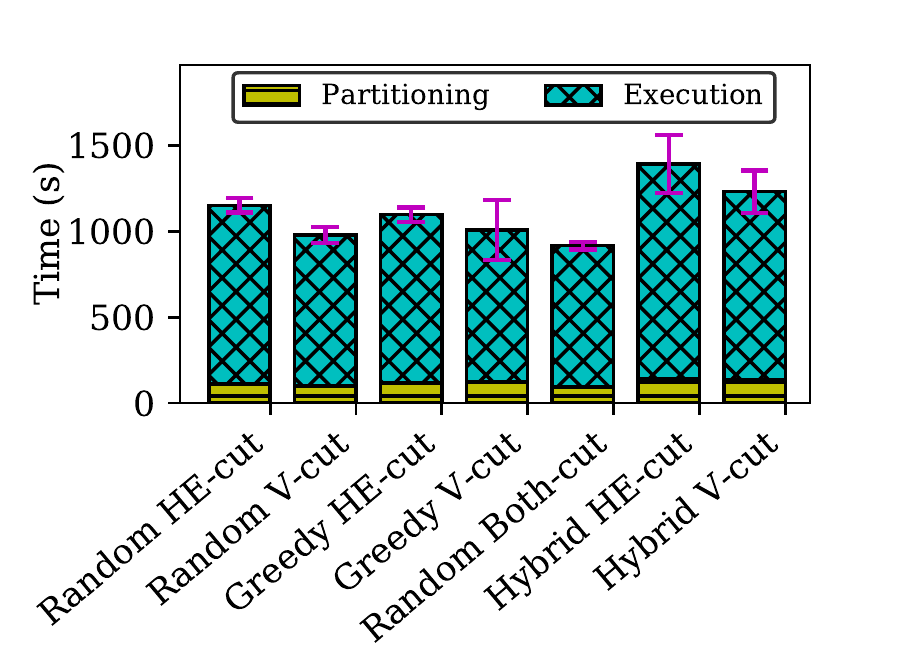}
		\label{fig:partitioning-sp-orkut}
	}
	\caption{Shortest Paths: Partitioning and execution time using several
		partitioning algorithms in \mesh{}.\label{fig:partitioning-sp}}
\end{figure*}

Next, we evaluate the partitioning policies described in
Section~\ref{sec:implementation}.
Due to space constraints, we omit results for the Apache dataset here.
Figure~\ref{fig:partitioning-lp} shows both partitioning time and subsequent
execution time for the Label Propagation algorithm for each of these policies
for the dblp, Friendster, and Orkut datasets.
Figures~\ref{fig:partitioning-pr}, 
\ref{fig:partitioning-prEntropy}, and 
\ref{fig:partitioning-sp}
repeat these experiments for the PageRank,
PageRank-Entropy, and Shortest Paths 
algorithms respectively.

We see that the choice of 
the best partitioning algorithm depends on the data.
One possible data characteristic having an impact could be the relative number of vertices and hyperedges in the hypergraph.
First considering Figures~\ref{fig:partitioning-lp} (Label Propagation) 
and \ref{fig:partitioning-pr} (PageRank),
we see that the greedy hyperedge-cut algorithm is the best for the
Friendster hypergraph (Figures \ref{fig:partitioning-lp-friendster} and \ref{fig:partitioning-pr-friendster}), where vertices outnumber hyperedges.
Here, cutting hyperedges while partitioning the larger set of vertices might lead to better computational load  balancing.
On the other hand,
for the Orkut hypergraph  (Figures \ref{fig:partitioning-lp-orkut} and \ref{fig:partitioning-pr-orkut}),
where hyperedges outnumber vertices,
we see that while the vertex-cut algorithms seem to perform better than the corresponding 
hyperedge-cut variants, a Random Both-cut algorithm is the best.
This suggests that cutting vertices is better than cutting hyperedges, but that cutting both sets may lead to even better load balancing.
For dblp (Figures \ref{fig:partitioning-lp-dblp} and \ref{fig:partitioning-pr-dblp}), we see a much less pronounced difference between vertex-cut and
hyperedge-cut algorithms, as the number of hyperedges and vertices in this
dataset are roughly equal.

Figures~\ref{fig:partitioning-prEntropy} and \ref{fig:partitioning-sp} shows the results for the PageRank-Entropy and Shortest Path algorithms, and their trends are very similar to those of the PageRank and Label Propagation algorithms,
including
the best partitioning strategy for each dataset.
Note that the execution times of Shortest Path algorithm shown in Figure~\ref{fig:partitioning-sp} are smaller than those of other algorithms because it terminates when messages are passed through the maximum distance between any two vertices, i.e., the diameter of the graph, whereas the other algorithms run more iterations until the values of vertices and hyperedges are converged or exceed the maximum number of iterations  (30 for our experiments). 

These results show that no one partitioning algorithm dominates all others in all cases.
The best choice depends on the characteristics of the hypergraph.
For instance, holistically partitioning the hypergraph,
as done by the Greedy vertex-cut and hyperedge-cut
algorithms, can be beneficial in some cases, while
cutting both hyperedges and vertices can be effective in others.
A promising next step is to develop a combined algorithm that partitions
holistically as the Greedy algorithms do, while differentiating between
hyperedges and vertices as the Random Both-cut and Hybrid algorithms do.

These results also show the value of the flexibility provided by \mesh{},
where the choice of an appropriate partitioning algorithm can be based on data and application characteristics.  
Note that the vertex-to-hyperedge ratio is only one data characteristic that may be impacting the performance.
Identifying all the relevant characteristics and their impact, and automatically making the design choices is an area of future work.

\subsection{Scaling}

Our next set of experiments examine the scaling of \mesh{} based on available computing resources (size of the cluster and size of the Spark workers). 
These experiments are carried out on the 65-node Amazon AWS testbed.
We show results for {\em strong scaling}, i.e., keeping the total dataset size the same, while changing the number of nodes and the number of Spark workers in the cluster.
We execute Label Propagation algorithm for hybrid vertex-cut partitioning strategies and for the largest dataset, Orkut, with up to 65 amazon AWS m4.2xlarge instances, and measured the partitioning and execution times. We omit results for other smaller datasets due to space constraints.

\begin{figure*}[htpb]
	\centering
	\subfigure[Varying number of workers (Spark cores) on 16-node cluster, partitioning and execution time (10 iterations, Orkut)]{
		\includegraphics[width=5.7cm]{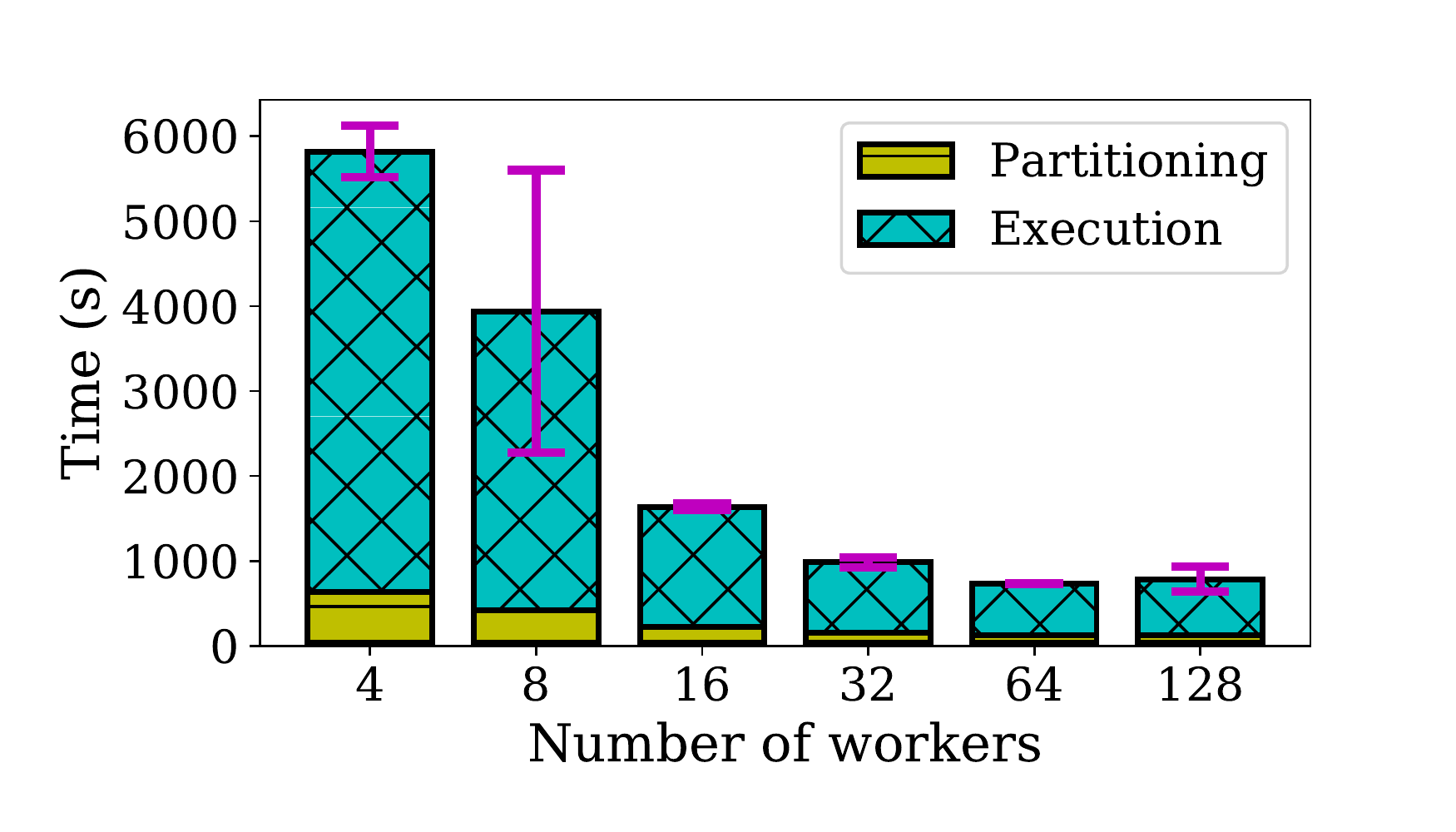}
		\label{fig:16-node-cluster-spark-cores}
	}
	\hspace{2cm}
	\subfigure[Varying the size of cluster, partitioning and execution time (10 iterations, Orkut).]{
		\includegraphics[width=5.7cm]{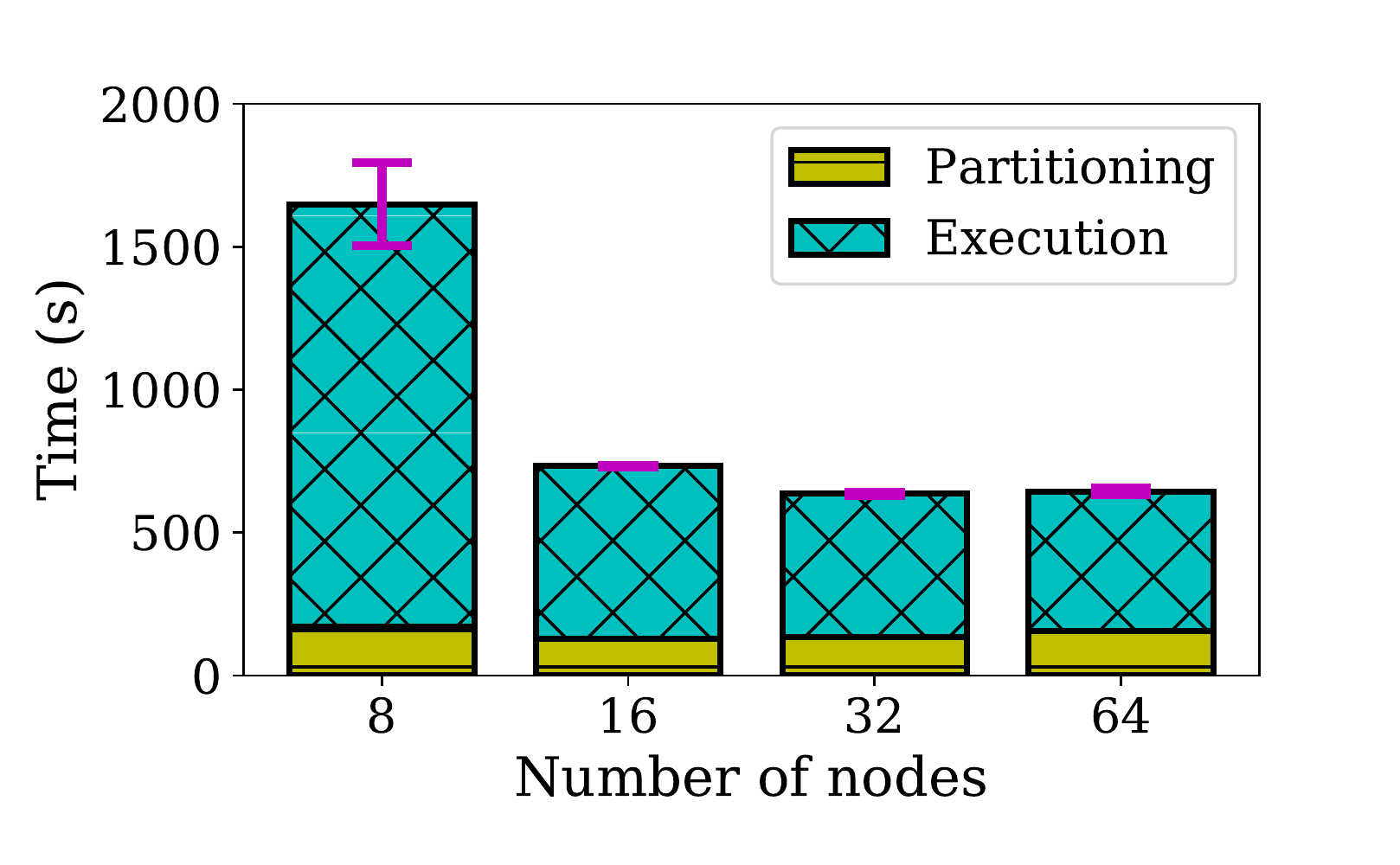}
		\label{fig:mesh_larger_scaling}
	}
	\vspace{-0.2cm}
	\caption{\mesh{} system scalability.\label{fig:scalability}}
	\vspace{-0.5cm}
\end{figure*}

Figure~\ref{fig:16-node-cluster-spark-cores} shows the scaling results for the Orkut dataset by increasing the number of Spark workers, while keeping the number of nodes fixed. 
\mesh{} is evaluated in a cluster with 1 master and 16 slaves (8 cores per node) and number of workers from ranging from 4 to 128.
As the number of workers increases, both partitioning and execution time decrease, and we get diminishing returns in performance improvement as we saturate the physical cores of the cluster nodes.

Figure~\ref{fig:mesh_larger_scaling} shows the scaling results for the Orkut dataset by increasing the number of nodes (number of slaves of a cluster) from 8 to 64 nodes. For example, 64-node cluster indicates 64 m4.2xlarge instances used for slaves and 1 m4.2xlarge for the master of the cluster. The partitioning and execution times for each cluster size are based on the best Spark cores setting for that cluster size. The figure shows the execution time decreases as the computing resources increase from 8 to 64 nodes. However, the partitioning time remains relatively constant even with larger set of computing resources, because a minimum amount of time is necessary for the completion of a partitioning task regardless of the cluster size.

% scaling-bar-graphs
\begin{figure*}[htbp]
	\centering
	\subfigure[dblp]{
		\includegraphics[width=0.60\columnwidth]{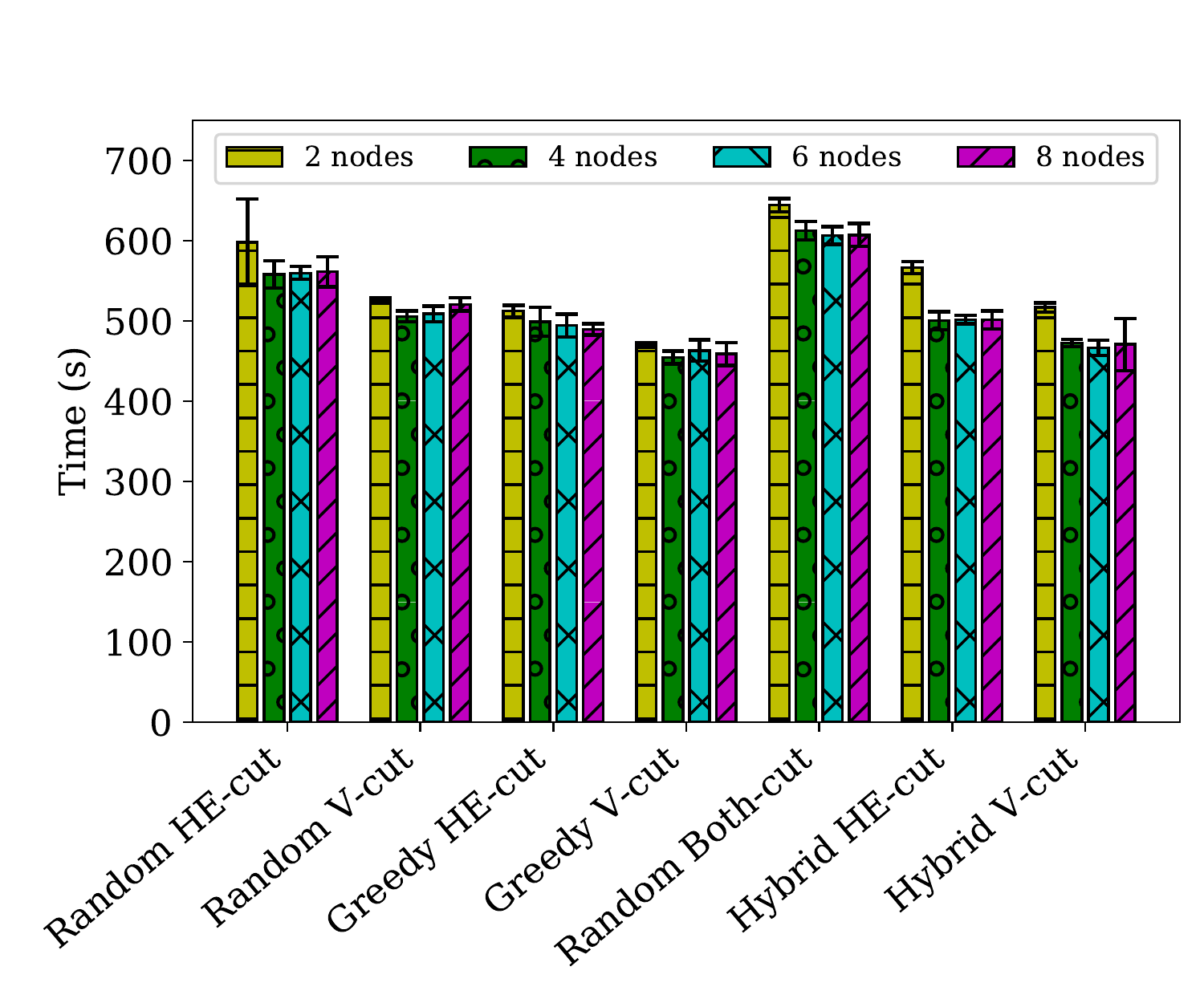}
		\label{fig:scaling_bar_dblp}
	}
	\subfigure[Friendster]{
		\includegraphics[width=0.60\columnwidth]{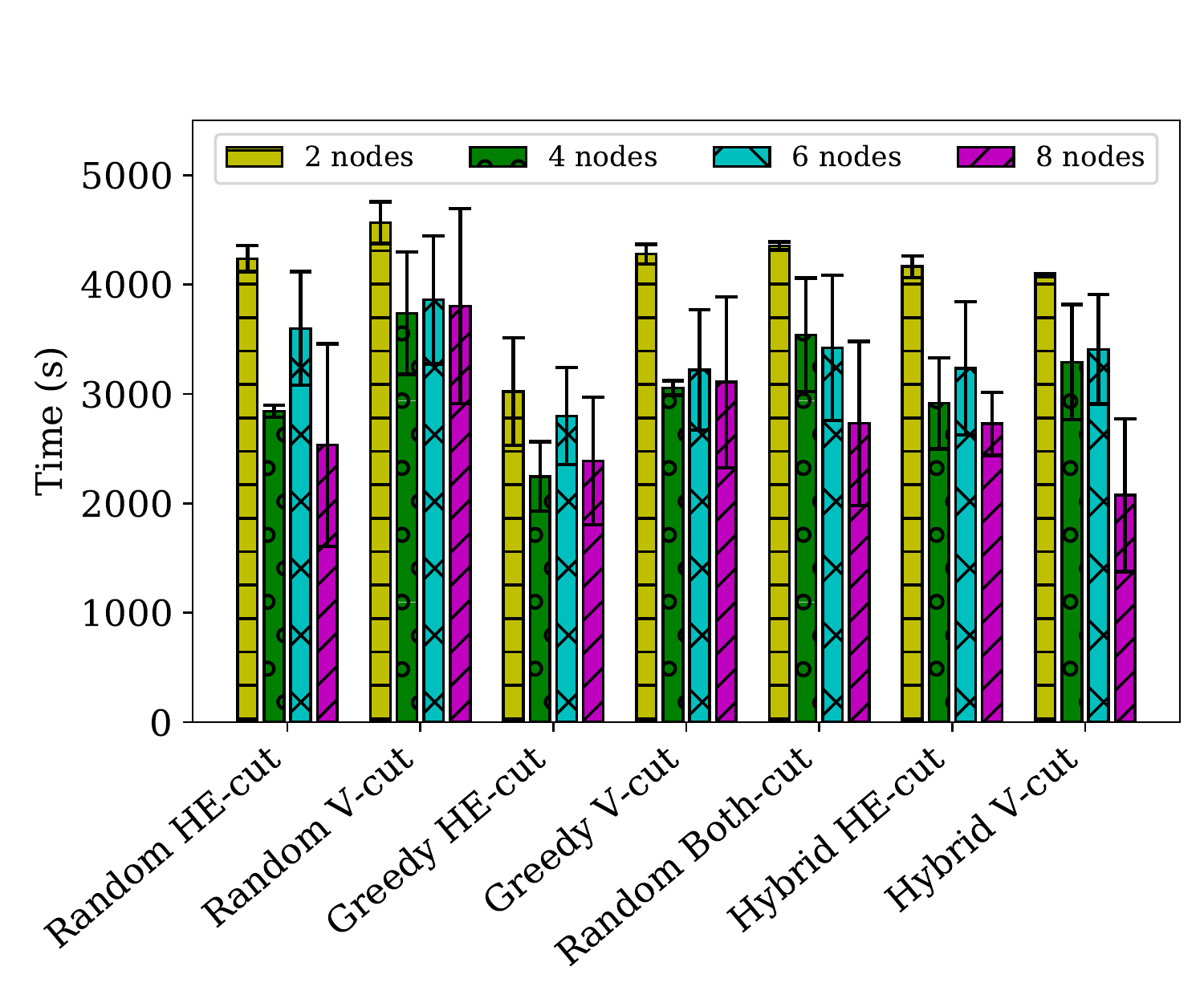}
		\label{fig:scaling_bar_friendster}
	}
	\subfigure[Orkut]{
		\includegraphics[width=0.60\columnwidth]{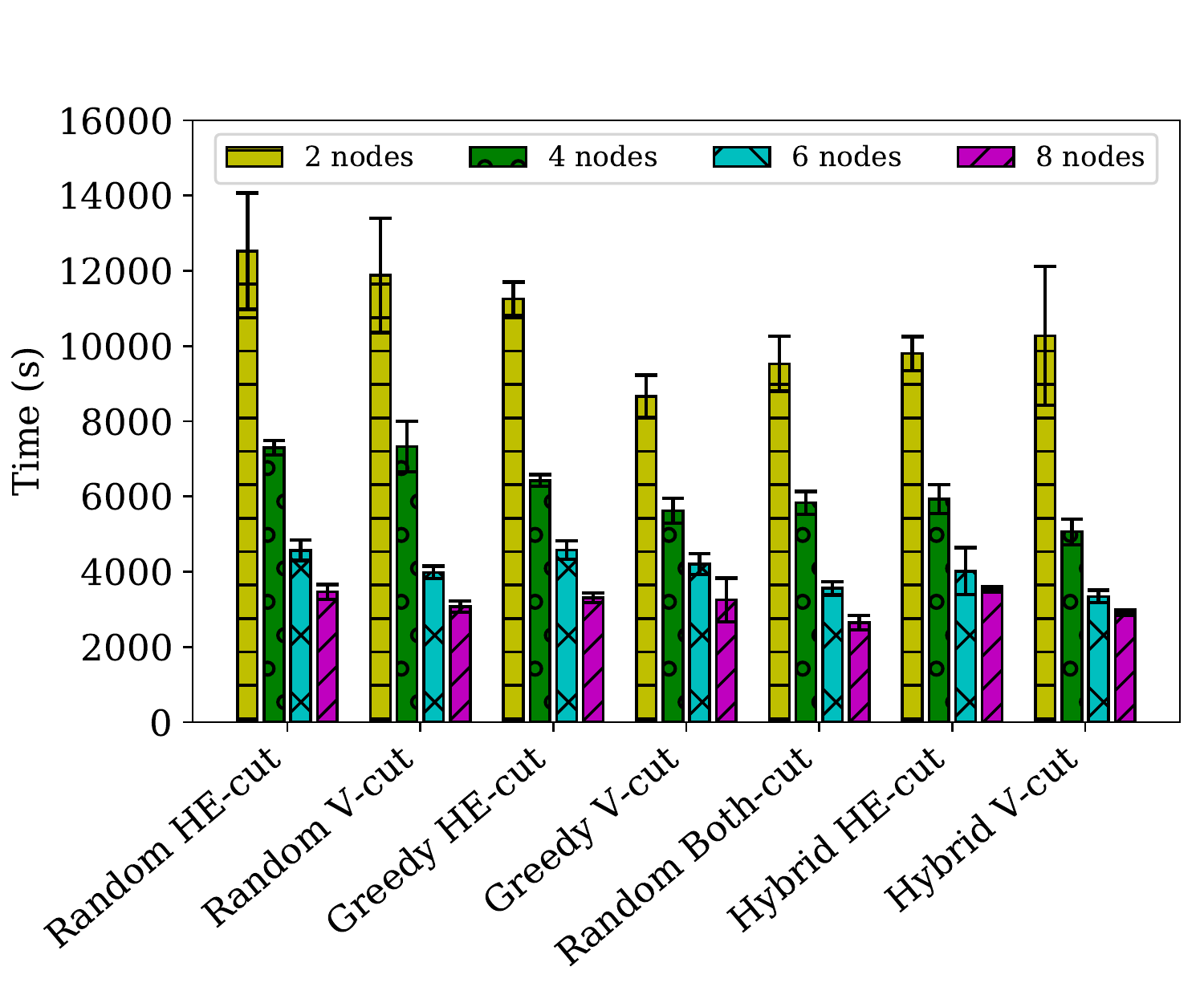}
		\label{fig:scaling_bar_orkut}
	}
	\vspace{-0.2cm}
	\caption{Partitioning and Label Propagation execution time for all partitioning strategies in various size of \mesh{} clusters.\label{fig:scaling_bar}}
	\vspace{-0.3cm}
\end{figure*}

\begin{figure}[htbp]
	\centering
	\includegraphics[width=0.6\columnwidth]{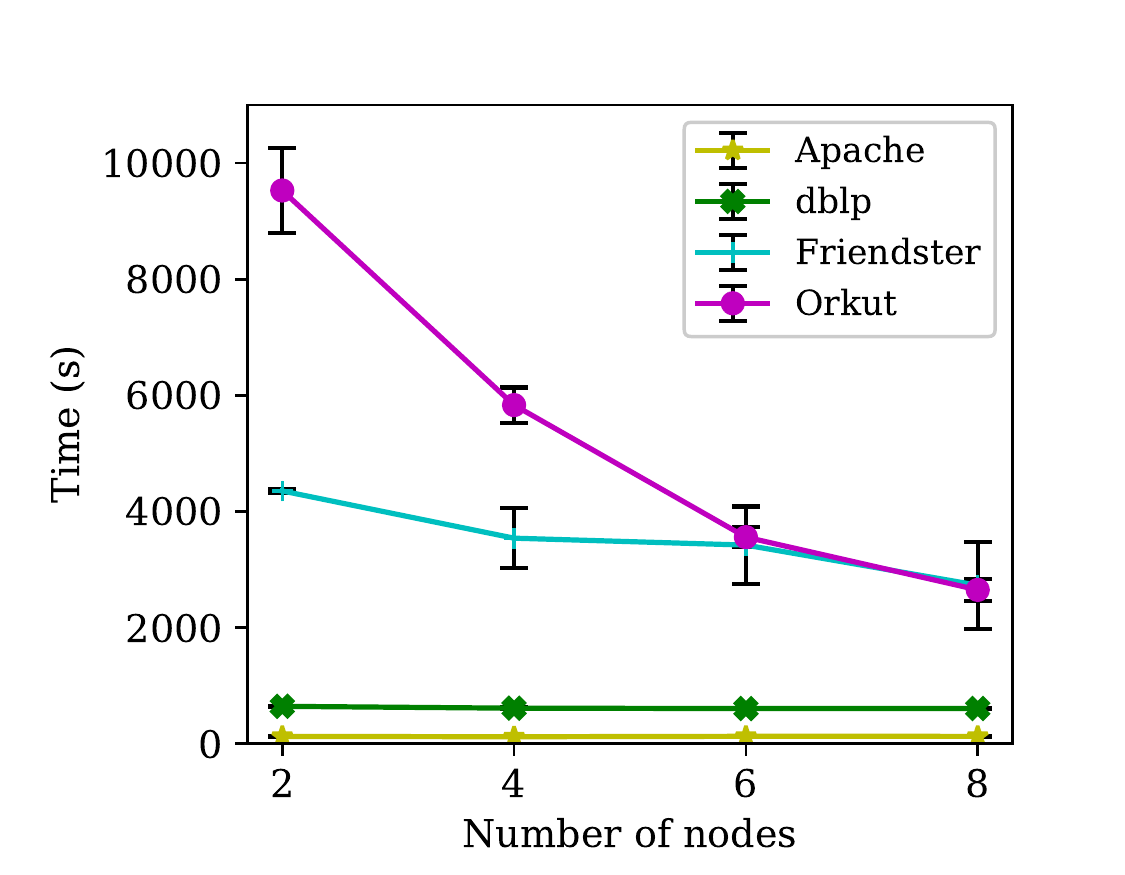}
	\caption{Execution time for the Label Propagation algorithm on various size datasets with Random Both-cut partitioning strategy.
	\label{fig:scaling_2d}}
	\vspace{-0.5cm}
\end{figure}

Figure~\ref{fig:scaling_2d} shows the scaling results for the four datasets using the Random Both-cut partitioning strategy and Figure~\ref{fig:scaling_bar} shows detailed results for all partitioning strategies (Apache is omitted due to space constraints).
We make the following observations from these results.
First, as seen from the figures, the execution times for all datasets either decreases or flattens out as the size of the cluster increases. The bigger the size of the dataset is, the greater the performance benefit.
For instance, in Figure~\ref{fig:scaling_bar_orkut}, for Orkut, the execution time keeps decreasing as the computing resources increase from 2 to 8 nodes, indicating a computational resource bottleneck at smaller cluster sizes.
However, if the computing resources are sufficient to deal with a given dataset (e.g., Apache and dblp), the performance improvement obtained by running on a larger cluster becomes insignificant. For example, Figure~\ref{fig:scaling_bar}(a) shows that for dblp, there is only a slight performance improvement when the size of the cluster increases from two to four nodes for all partition strategies. This means that a 2-node cluster is sufficient for the dblp dataset.

Figure~\ref{fig:scaling_2d} also shows that, although the Friendster and Orkut datasets have different input data sizes, their execution times are saturated around 3000 seconds with 8 nodes. 
Orkut dataset contains two times more the total numbers of vertices and hyperedges and 4.5 times more bipartite edges than Friendster dataset, but the maximum cardinality of Orkut is similar to that of Friendster (about 9,000 from Table~\ref{tab:datasets}).
Once the cluster is large enough that computational resources are sufficient, the network I/O becomes a bottleneck. Since the messages sent from one host to another are merged before they are sent out, the maximum cardinality influences the message sizes and hence, the total network traffic, leading to a similar performance for both Orkut and Friendster at larger cluster sizes.

\subsection{Comparison with HyperX}

To evaluate the overall performance, simplicity and flexibility of \mesh{}, we compare it against
HyperX~\cite{huang2015}, a hypergraph processing system that is also
built on top of Apache Spark.
Unlike \mesh{}, which builds on top of GraphX, HyperX implements a hypergraph
layer---heavily inspired by GraphX---directly on top of Spark.
While the HyperX implementation is optimized for hypergraph execution, our implementation relies on the GraphX optimizations designed for graph execution.
Here, we evaluate the performance tradeoff given the simplicity and flexibility of our API and implementation.

\begin{minipage}{\linewidth}
	\bigskip
	\centering
	\small
	\captionof{table}{\mesh{} vs. HyperX (Scala Lines of Code)}
	\label{tab:loc_comparison}
	\begin{tabular}{ | l | c | c | }
		\hline
		LOC & \mesh{} & HyperX \\ \hline
		System core & 630 & 2,620  \\ \hline
		Partition core & 30 & 1,295 \\ \hline
		Partition algorithm & 5 - 40 & 10 - 60 \\ \hline
		Total system & 795 & 4,050 \\ \hline
		Applications & LP: 35, RW: 40 & LP: 50, RW: 75 \\ \hline
	\end{tabular}
	\bigskip
\end{minipage}

\noindent{\bf Simplicity and flexibility comparison.} %To compare their simplicity and flexibility, 
Table~\ref{tab:loc_comparison} shows the quantitative difference in the implementations of \mesh{} and HyperX in terms of the lines of code (LOC) required for each system. System core code corresponds to the core system functionality such as handling hypergraph construction, processing, etc. 
Partition core code is specifically related to basic partitioning features. 
Partition algorithm indicates the range of lines of codes needed to implement a particular partitioning algorithm (e.g., Hybrid Vertex-cut) based on the partition and system cores. 
Total system is the sum of LOCs for core and partition algorithms. 

We see that HyperX requires 5 times more LOCs, compared to \mesh{} since it directly builds on top of Spark. In terms of partition modules, HyperX needs 43 times more LOCs to partition a hypergraph in their system than \mesh{}, as \mesh{} is able to take advantage of the basic partitioning functionality provided by GraphX. 
Besides, HyperX requires slightly more LOCs to implement a particular partitioning algorithm compared to \mesh{}. 
This shows that \mesh{} is much simpler and more flexible compared to HyperX, allowing various system features and partitioning algorithms for distributed hypergraph computation. 
In addition, in terms of ease of use, 
\mesh{} applications can be implemented with fewer LOCs as compared to HyperX.
For example, as shown in Table~\ref{tab:loc_comparison}, Label Propagation (LP) and Random Walk (RW) require 35 and 40 LOCs on \mesh{}, compared to 50 and 75 for HyperX. 

\noindent{\bf Performance comparison.} We compare the performance of these two systems using a Label Propagation algorithm, specifically
Listing~\ref{listing:lp} for \mesh{}, and the provided example implementation for
HyperX.\footnote{We modify the implementation in HyperX to compute over undirected
	hypergraphs.}

\begin{figure}[htbp]
  \centering
  \includegraphics[width=0.85\columnwidth]{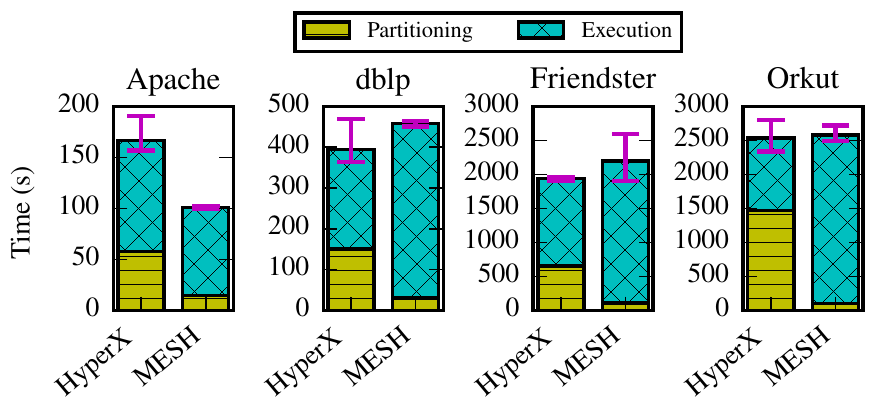}
  \caption{Partitioning and Label Propagation execution time for \mesh{} (using
    the best partitioning algorithm) and HyperX.\label{fig:hyperx-lp}}
  \vspace{-0.1cm}
\end{figure}

Figure~\ref{fig:hyperx-lp} shows the partitioning and Label Propagation
execution times (for 30 iterations) on the local 8-node cluster for HyperX and \mesh{} (using the best
partitioning policies for the given dataset).
Unlike \mesh{}, HyperX uses an iterative partitioning algorithm (10 iterations in
our experiments, based on the HyperX experiments~\cite{huang2015}), 
leading to much higher partitioning times and comparable total running times.

In terms of performance, these results show the efficacy of \mesh{}, 
which achieves comparable performance to HyperX,
despite lacking several low-level optimizations.
An additional qualitative benefit of \mesh{} is its flexibility:
hypergraphs are diverse, and \mesh{} provides a simple interface that allows implementing different partitioning
policies easily, as shown above.

From a higher level, our results suggest that high performance need not be at
odds with a simple and flexible implementation.
In fact, by layering on top of GraphX and leveraging its maturity and ongoing
development, we can expect to reap the benefits of ongoing optimization.
Backporting future optimizations to HyperX, on the other hand, would require
significant engineering effort.

     % label: sec:evaluation
\section{Related Work}
\label{sec:related}

%\noindent{\bf Graph Processing Systems:}
\paragraph{Graph Processing Systems}
There has been a flurry of research on graph computing systems in recent
years~\cite{
  malewicz2010,
  gonzalez2012,
  %bu2014, 
  %cheng2012,
  %kyrola2012,
  %low2012,
  %low2010,
  %satish2014,
  salihoglu2013},
and along with it, a great deal of work on
performance evaluation and optimization~\cite{han2014_a, salihoglu2014, zhang2011}.
Key among these systems, Pregel~\cite{malewicz2010} introduced the ``think like
a vertex'' model.
GraphX~\cite{gonzalez2014}, built upon Apache Spark~\cite{zaharia2012},
adopted a similar model while
inheriting the scalability and fault tolerance of Spark's Resilient Distributed
Datasets (RDD).
%GraphLab~\cite{gonzalez2012, low2010, low2012}
GraphLab~\cite{low2010}
provided a more fine-grained
interface along with support for asynchronous computation.

These systems provide scalability, and their interfaces are easy to use in
the graph computing context.
Our \mesh{} API can be viewed as an extension of the ``think like a vertex''
model.
Although we have discussed challenges in implementing \mesh{} on top of a graph
processing system in general, and GraphX in particular, there is no fundamental
requirement that \mesh{} run on top of a specialized platform.
For example, \mesh{} could be implemented on top of a general-purpose relational
database management system (RDBMS)~\cite{fan2015}.
GraphX, however, is particularly compelling due to the popularity and growth of
Spark.
Further, by facilitating diverse views of the same underlying data---e.g.,
collection-oriented, graph-oriented, tabular~\cite{armbrust2015}---building on
top of Spark allows easier integration in broader data processing pipelines.

%\noindent{\bf Graph and Hypergraph Partitioning:}
\paragraph{Graph and Hypergraph Partitioning}
Graph Partitioning is a significant research topic in its own right.
In the high-performance computing context, metis~\cite{karypis1996} provides
very effective graph partitioning, and has open-source implementations
for both single-node and distributed deployment.
Its hMetis~\cite{karypis1996} cousin partitions hypergraphs, but no
distributed implementation yet exists.
The Zoltan toolkit from Sandia National Laboratories~\cite{zoltan} includes
a parallel hypergraph partitioner~\cite{zoltan-partitioning} that cuts both
vertices and hyperedges.

In the distributed systems context, PowerGraph~\cite{gonzalez2012} targets
natural (e.g., social) graphs by cutting vertices rather than edges.
While this is effective for natural graphs, hypergraphs require different
approaches.
Chen et al. have proposed novel algorithms for bipartite
graphs~\cite{gonzalez2014} and skewed graphs~\cite{chen2015}, which we have
used as the basis for our Greedy and Hybrid algorithms, respectively.
While these are already effective algorithms, there remains opportunity
to combine holistic and differentiated approaches to improve
hypergraph partitioning.

%\noindent{\bf Hypergraph Processing:}
\paragraph{Hypergraph Processing}
Hypergraphs have been studied for
%decades~\cite{estrada2005,berge1976,berge1989}
decades~\cite{estrada2005,berge1976}
and have been applied in many
settings, ranging from bioinformatics~\cite{gallagher2014} to VLSI
%design~\cite{alpert1995,karypis1999}
design~\cite{karypis1999}
to database optimization~\cite{liu1996}.
Social networks have generally been modeled using simple graphs, but hypergraph
variants of popular graph algorithms (e.g., centrality
estimation~\cite{bonacich2004,roy2015}, shortest paths~\cite{gao2012}) have
been developed in recent years.
HyperX~\cite{huang2015} builds a hypergraph processing system on top of Spark,
but does so by modifying GraphX rather than building on top of GraphX.
Unlike HyperX, \mesh{} does not make any static assumptions about the data
characteristics, and instead provides the flexibility necessary to choose an
appropriate representation and partitioning algorithm at runtime based on data
and application characteristics.

% In prior work~\cite{heintz2014}, we have explored other issues surrounding
% hypergraph computing, such as application-level modeling considerations and the
% need for characterization of real-world hypergraphs.
% In this paper, on the other hand, we have proposed a concrete programming
% interface and explored some of the specific challenges that arise when
% implementing this interface on existing systems.
% Hypergraphs have been studied for decades~\cite{berge1976,berge1989} and have
% been applied in diverse applications such as VLSI design~\cite{karypis1999}.
% Our hope is that scalable hypergraph computing \emph{systems} will enable and
% encourage the development of novel and useful hypergraph \emph{algorithms}.
        % label: sec:related
\section{Conclusion}
We presented \mesh{}, a flexible distributed framework for
scalable hypergraph processing based on a graph processing system.
\mesh{} provides an easy-to-use and expressive API that naturally extends the
``think like a vertex'' model common to many popular graph processing systems.
We used our system to explore the key challenges in implementing a
hypergraph processing system on top of a graph processing system: 
how to partition the hypergraph representation to allow
distributed computation.
\mesh{} provides flexibility to implement different design choices, and
by implementing \mesh{} on top of the popular GraphX framework, we have leveraged
the maturity and ongoing development of the Spark ecosystem and kept our
implementation simple.
% for different data characteristics.
Our experiments with multiple real-world datasets and algorithms on a local cluster as well as an AWS testbed demonstrated that this
flexibility does not come at the expense of performance, as even our
unoptimized prototype performs comparably to HyperX.

\begin{comment}
As part of future work, we would like to provide automated runtime selection
of the best choice of implementation choices based on data and
application characteristics.
Other extensions include support for directed hyperedges and asynchronous
computation.
\end{comment}

%-------------------------------------------------------------------------------
\section*{Acknowledgments}
%-------------------------------------------------------------------------------

This work is supported in part by NSF grant III-1422802.
     % label: sec:conclusion
\bibliographystyle{IEEEtran}
\bibliography{refbib.bib}

\begin{thebibliography}{10}
\providecommand{\url}[1]{#1}
\csname url@samestyle\endcsname
\providecommand{\newblock}{\relax}
\providecommand{\bibinfo}[2]{#2}
\providecommand{\BIBentrySTDinterwordspacing}{\spaceskip=0pt\relax}
\providecommand{\BIBentryALTinterwordstretchfactor}{4}
\providecommand{\BIBentryALTinterwordspacing}{\spaceskip=\fontdimen2\font plus
\BIBentryALTinterwordstretchfactor\fontdimen3\font minus
  \fontdimen4\font\relax}
\providecommand{\BIBforeignlanguage}[2]{{%
\expandafter\ifx\csname l@#1\endcsname\relax
\typeout{** WARNING: IEEEtran.bst: No hyphenation pattern has been}%
\typeout{** loaded for the language `#1'. Using the pattern for}%
\typeout{** the default language instead.}%
\else
\language=\csname l@#1\endcsname
\fi
#2}}
\providecommand{\BIBdecl}{\relax}
\BIBdecl

\bibitem{gonzalez2014}
J.~E. Gonzalez, R.~S. Xin, A.~Dave, D.~Crankshaw, M.~J. Franklin, and
  I.~Stoica, ``Graph{X}: Graph processing in a distributed dataflow
  framework,'' in \emph{Proc. of OSDI}, 2014, pp. 599--613.

\bibitem{malewicz2010}
G.~Malewicz, M.~H. Austern, A.~J. Bik, J.~C. Dehnert, I.~Horn, N.~Leiser, and
  G.~Czajkowski, ``Pregel: A system for large-scale graph processing,'' in
  \emph{Proc. of SIGMOD ICMD}, 2010, pp. 135--146.

\bibitem{nguyen2013}
D.~Nguyen, A.~Lenharth, and K.~Pingali, ``A lightweight infrastructure for
  graph analytics,'' in \emph{Proc. of SOSP}, 2013, pp. 456--471.

\bibitem{lazer2006}
D.~Lazer, A.~Pentland, L.~Adamic, S.~Aral, A.-L. Barabási, D.~Brewer,
  N.~Christakis, N.~Contractor, J.~Fowler, M.~Gutmann, T.~Jebara, G.~King,
  M.~Macy, D.~Roy, and M.~Van~Alstyne, ``Computational social science,''
  \emph{Science}, vol. 323, no. 5915, pp. 721--723, 2009.

\bibitem{estrada2005}
E.~Estrada and J.~Rodriguez-Velazquez, ``Complex networks as hypergraphs,''
  \emph{Arxiv preprint physics/0505137}, 2005.

\bibitem{sharma2017weighted}
A.~Sharma, T.~J. Moore, A.~Swami, and J.~Srivastava, ``Weighted simplicial
  complex: A novel approach for predicting small group evolution,'' in
  \emph{Pacific-Asia Conference on Knowledge Discovery and Data Mining}.\hskip
  1em plus 0.5em minus 0.4em\relax Springer, 2017, pp. 511--523.

\bibitem{gallagher2014}
S.~R. Gallagher, M.~Dombrower, and D.~S. Goldberg, ``Using 2-node hypergraph
  clustering coefficients to analyze disease-gene networks,'' in \emph{Proc. of
  BCB}, 2014, pp. 647--648.

\bibitem{berge1976}
C.~Berge, \emph{Graphs and hypergraphs}.\hskip 1em plus 0.5em minus 0.4em\relax
  Elsevier, 1976, vol.~6.

\bibitem{huang2015}
J.~Huang, R.~Zhang, and J.~X. Yu, ``Scalable hypergraph learning and
  processing,'' in \emph{Proc. of ICDM}, Nov 2015, pp. 775--780.

\bibitem{zaharia2012}
M.~Zaharia, M.~Chowdhury, T.~Das, A.~Dave, J.~Ma, M.~McCauley, M.~J. Franklin,
  S.~Shenker, and I.~Stoica, ``Resilient distributed datasets: A fault-tolerant
  abstraction for in-memory cluster computing,'' in \emph{Proc. of NSDI}, 2012.

\bibitem{gonzalez2012}
J.~E. Gonzalez, Y.~Low, H.~Gu, D.~Bickson, and C.~Guestrin, ``Power{G}raph:
  Distributed graph-parallel computation on natural graphs,'' in \emph{Proc. of
  OSDI}, 2012, pp. 17--30.

\bibitem{page1999}
L.~Page, S.~Brin, R.~Motwani, and T.~Winograd, ``The {P}age{R}ank citation
  ranking: Bringing order to the web,'' Stanford InfoLab, Tech. Rep., 1999.

\bibitem{raghavan2007}
U.~N. Raghavan, R.~Albert, and S.~Kumara, ``Near linear time algorithm to
  detect community structures in large-scale networks,'' \emph{Phys. Rev. E},
  vol.~76, p. 036106, Sep 2007.

\bibitem{roy2015}
S.~Roy and B.~Ravindran, ``Measuring network centrality using hypergraphs,'' in
  \emph{Proc. of CoDS}, 2015, pp. 59--68.

\bibitem{chen2015}
R.~Chen, J.~Shi, Y.~Chen, and H.~Chen, ``Power{L}yra: Differentiated graph
  computation and partitioning on skewed graphs,'' in \emph{Proc. of EuroSys},
  2015, pp. 1:1--1:15.

\bibitem{chen2014}
R.~Chen, J.~Shi, B.~Zang, and H.~Guan, ``Bipartite-oriented distributed graph
  partitioning for big learning,'' in \emph{Proc. of APSys}, 2014, pp.
  14:1--14:7.

\bibitem{snapnets}
J.~Leskovec and A.~Krevl, ``{SNAP Datasets}: {Stanford} large network dataset
  collection,'' \url{http://snap.stanford.edu/data}, Jun. 2014.

\bibitem{apache-ssf}
``The apache software foundation,'' \url{http://www.apache.org/}, 2017.

\bibitem{salihoglu2013}
S.~Salihoglu and J.~Widom, ``{GPS}: A graph processing system,'' in \emph{Proc.
  of SSDBM}, 2013, pp. 22:1--22:12.

\bibitem{han2014_a}
M.~Han, K.~Daudjee, K.~Ammar, M.~T. \"{O}zsu, X.~Wang, and T.~Jin, ``An
  experimental comparison of pregel-like graph processing systems,''
  \emph{Proc. VLDB Endow.}, vol.~7, no.~12, pp. 1047--1058, Aug. 2014.

\bibitem{salihoglu2014}
S.~Salihoglu and J.~Widom, ``Optimizing graph algorithms on pregel-like
  systems,'' \emph{Proc. VLDB Endow.}, vol.~7, no.~7, pp. 577--588, Mar. 2014.

\bibitem{zhang2011}
Y.~Zhang, Q.~Gao, L.~Gao, and C.~Wang, ``{P}r{I}ter: A distributed framework
  for prioritized iterative computations,'' in \emph{Proc. of SOCC}, 2011, pp.
  13:1--13:14.

\bibitem{low2010}
Y.~Low, J.~Gonzalez, A.~Kyrola, D.~Bickson, C.~Guestrin, and J.~M. Hellerstein,
  ``{G}raph{L}ab: A new parallel framework for machine learning,'' in
  \emph{Proc. of UAI}, July 2010.

\bibitem{fan2015}
J.~Fan, A.~G.~S. Raj, and J.~M. Patel, ``The case against specialized graph
  analytics engines,'' in \emph{Proc. of CIDR}, 2015.

\bibitem{armbrust2015}
M.~Armbrust, R.~S. Xin, C.~Lian, Y.~Huai, D.~Liu, J.~K. Bradley, X.~Meng,
  T.~Kaftan, M.~J. Franklin, A.~Ghodsi, and M.~Zaharia, ``Spark {SQL}:
  Relational data processing in spark,'' in \emph{Proc. of SIGMOD}, 2015, pp.
  1383--1394.

\bibitem{karypis1996}
G.~Karypis and V.~Kumar, ``Parallel multilevel k-way partitioning scheme for
  irregular graphs,'' in \emph{Proc. of SC}, 1996.

\bibitem{zoltan}
``Zoltan: Data management services for parallel applications,''
  \url{http://www.cs.sandia.gov/Zoltan/}, 2016.

\bibitem{zoltan-partitioning}
K.~D. Devine, E.~G. Boman, R.~T. Heaphy, R.~H. Bisseling, and U.~V. Catalyurek,
  ``Parallel hypergraph partitioning for scientific computing,'' in \emph{Proc.
  of IPDPS}, April 2006.

\bibitem{karypis1999}
G.~Karypis, R.~Aggarwal, V.~Kumar, and S.~Shekhar, ``Multilevel hypergraph
  partitioning: applications in {VLSI} domain,'' \emph{IEEE Transactions on
  VLSI Systems}, vol.~7, no.~1, pp. 69--79, March 1999.

\bibitem{liu1996}
D.-R. Liu and S.~Shekhar, ``Partitioning similarity graphs: A framework for
  declustering problems,'' \emph{Information Systems}, vol.~21, no.~6, pp.
  475--496, 1996.

\bibitem{bonacich2004}
P.~Bonacich, A.~C. Holdren, and M.~Johnston, ``Hyper-edges and multidimensional
  centrality,'' \emph{Social Networkrks}, vol.~26, no.~3, pp. 189 -- 203, 2004.

\bibitem{gao2012}
J.~Gao, Q.~Zhao, W.~Ren, A.~Swami, R.~Ramanathan, and A.~Bar-Noy, ``Dynamic
  shortest path algorithms for hypergraphs,'' \emph{Arxiv preprint
  arXiv:1202.0082}, 2012.

\end{thebibliography}

\end{document}